\documentclass[aps,prapplied,reprint,superscriptaddress]{revtex4-1}
\usepackage[utf8]{inputenc}

\usepackage{graphicx}
\usepackage{hyperref}
\usepackage[mathlines]{lineno}
\usepackage{amsmath}
\usepackage{xfrac}
\usepackage{soul}
\usepackage{cancel}
\usepackage{xcolor}
\usepackage[nameinlink,capitalize]{cleveref}
\usepackage{placeins}

\draft 

\begin{document}


\title{Hybrid Quantum Interferometer in Bifurcation Mode as a Latching Quantum Readout} 

\author{Connor D.~Shelly}
\email[Corresponding author:~]{cshelly@oxfordquantumcircuits.com}
\altaffiliation{Present Address:~Oxford Quantum Circuits, 46 Woodstock Road, Oxford, OX2 6HT, United Kingdom}
\affiliation{Royal Holloway, University of London, Egham, TW20 0EX, United Kingdom}
\affiliation{National Physical Laboratory, Hampton Road, Teddington, TW11 0LW, United Kingdom}

\author{Christopher Checkley}
\affiliation{Royal Holloway, University of London, Egham, TW20 0EX, United Kingdom}

\author{Victor T.~Petrashov}
\affiliation{Royal Holloway, University of London, Egham, TW20 0EX, United Kingdom}

\date{\today}

\begin{abstract}
We have developed a new type of magnetometer consisting of a Hybrid Quantum Interference Device (HyQUID) that is set in a bi-stable state. We demonstrate its operation in a latching mode that can be employed to measure small changes in the applied flux. The device can be used to probe the flux state of a superconducting circuit using straightforward electrical resistance measurements, making it suitable as a simple qubit readout with low back-action.
\end{abstract}


\maketitle 

\section{Introduction}
The ability to detect extremely small changes in magnetic flux is of paramount importance in a number of applications. One area of particular interest is the readout of superconducting flux qubits. Here, it is necessary that the detector is not only able to measure extremely small changes in the flux generated by the qubit but also that it has minimal back-action so that it does not induce quantum decoherence.

The quantum state of the qubit can be determined by measurement of the magnetic flux in the system. Typically a superconducting quantum interference device (SQUID) has been used to achieve this \cite{Chiorescu_Science_2003}. 

Existing SQUID-based readout methods of the state of a qubit are unsatisfactory for a number of reasons. First, to produce a readout the SQUID is switched into a finite voltage state, a process that strongly disturbs both the qubit circuit and the SQUID itself. Bursts of non-equilibrium quasiparticles are created with energies exceeding the superconductor gap, thus `poisoning' the qubit circuit and leading to decoherence \cite{Martinis_PRL_2009}. Second, due to the AC Josephson effect the voltage across the SQUID produces a microwave voltage pulse that can drive neighbouring qubits into their excited states \cite{Mannik_PRL_2004}. Whilst being useful for proof-of-principle purposes, switching methods are unsuitable for simultaneous measurements of multiple qubits, or experiments in which the preservation of the qubit state after the measurement is required (for example, quantum non-demolition measurements \cite{Lupascu_PRL_2006}).


In an effort to overcome the above problems, an alternative readout device was developed - the Josephson Bifurcation Amplifier (JBA) \cite{Siddiqi_PRL_2004}. The JBA is essentially a nonlinear oscillator formed by a capacitively shunted Josephson junction. The JBA uses a dispersive measurement technique to avoid switching of the junction into a finite voltage state. The principle of the measurement technique is to drive the system with a sufficiently large rf excitation whilst measuring the plasma frequency response \cite{Lupascu_PRL_2006,Siddiqi_PRB_2006}. This high drive power causes the oscillator to enter a nonlinear regime in which a bistablity occurs \cite{Vijay_RevSciInst_2009}. As the plasma frequency response also varies with the critical current of the junction it is possible to use the JBA as a sensitive threshold detector. A particular advantage of this measurement technique is that the JBA will remain in the same state post measurement which allows the system to be employed as a latching readout of superconducting circuits such as flux qubits \cite{deGroot_APL_2010,Lupascu_NatPhys_2007}.

Although the JBA addresses a number of the shortcomings of SQUID-based measurements, there are still disadvantages to the method, namely the necessity for a large amplitude electromagnetic field in the system. The necessary coupling to the resonant cavity can also lead to photon-induced dephasing of the qubit \cite{Schuster_PRL_2007, Sears_dephasing_2012}. Finally, in order to properly drive the JBA a complicated and expensive arrangement of transmission lines, circulators, and rf electronics are required.

Hybrid Quantum Interference Devices (HyQUIDs) are hybrid mesoscopic devices that act as sensitive detectors of superconducting phase \cite{Petrashov_PRL_1995,Petrashov_PRL_2005,Shelly2016}. In this paper we show how a HyQUID can be set in a bi-stable state exhibiting a similar latching action to the JBA. This can be achieved without the need for the aforementioned complicated rf equipment - instead only the measurement of a quasi-dc voltage is needed to perform the HyQUID readout.

\section{The Hybrid Quantum Interference Device}
\subsection{HyQUID Operation Principles}

\begin{figure}
 \includegraphics[width=\columnwidth]{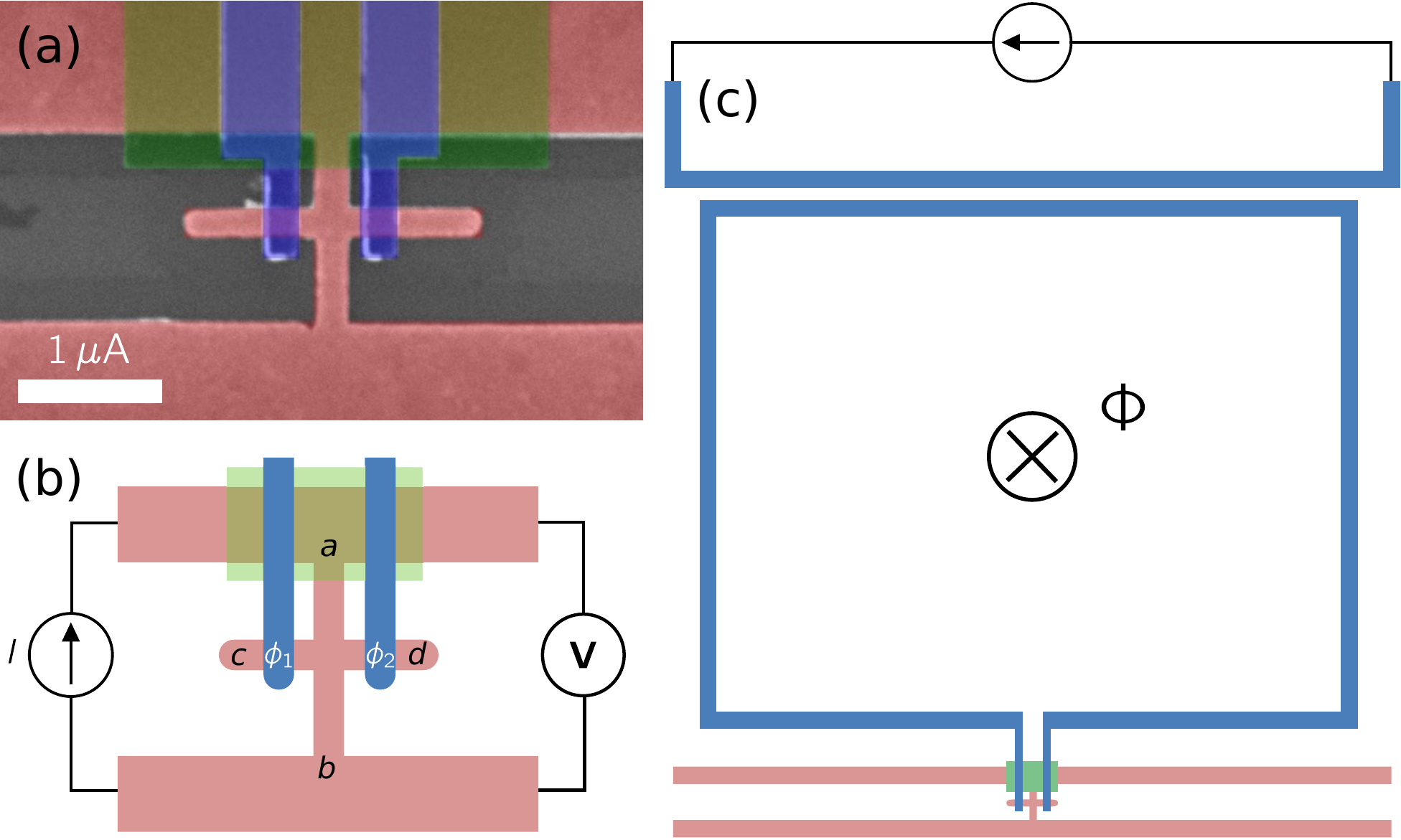}%
\caption{\label{fig:1}(a) Scanning electron micrograph of the HyQUID cross. The silver layer (normal metal) is coloured red, the aluminium layer (superconductor) is coloured blue, and the silicon-monoxide layer (insulating spacer) is coloured green. (b) Circuit schematic showing the HyQUID cross. One branch of the cross is connected to two normal metal reservoirs - a four-point measurement is used to monitor the resistance of the cross. The resistance of the normal branch oscillates as a function of the phase difference between the two superconductors $\phi=\phi_{1}-\phi_{2}$. (c) Schematic demonstrating the principle of operation of the HyQUID. The superconducting electrodes are joined to form a loop. A magnetic flux $\Phi$ can be applied to the loop using either an on-chip flux line (shown) or an external solenoid. The flux changes the phase difference across the interferometer and the resulting change in the HyQUID resistance is measured.}%
 \end{figure}

The HyQUID consists of a superconducting loop interrupted by a normal conductor weak link. A scanning electron micrograph and device schematics are shown in \cref{fig:1}. The normal cross of the HyQUID makes contact to normal electrodes at points \mbox{\textit{a} and \textit{b}} and to the superconducting loop at points \mbox{\textit{c} and \textit{d}}. The electrical resistance of the interferometer between points \textit{a} and \textit{b} oscillates as a function of the superconducting phase difference $\phi=\phi_{1}-\phi_{2}$ between points \textit{c} and \textit{d}, and is described by, 
\begin{equation}
\Delta R_{N} = \gamma(1-\cos(\phi))
\label{eq:andreev}
\end{equation}
where $\gamma$ is an amplitude factor controlled by properties of the system such as the quality of the SN interface and depends on the relation between the lengths of the normal conductor $L_{\mathrm{SNS}}$ connecting the superconductors and on characteristic length scales such as the electron phase breaking length $L_{\varphi}=\sqrt{D\tau_{\varphi}}$ and the coherence length $\xi_{\mathrm{N}}=\sqrt{\hbar D / 2\pi k_{\mathrm{B}}T}$, where $\tau_{\varphi}$ is the phase breaking time and $D$ is the diffusion coefficient.

\begin{figure}
 \includegraphics[width=\columnwidth]{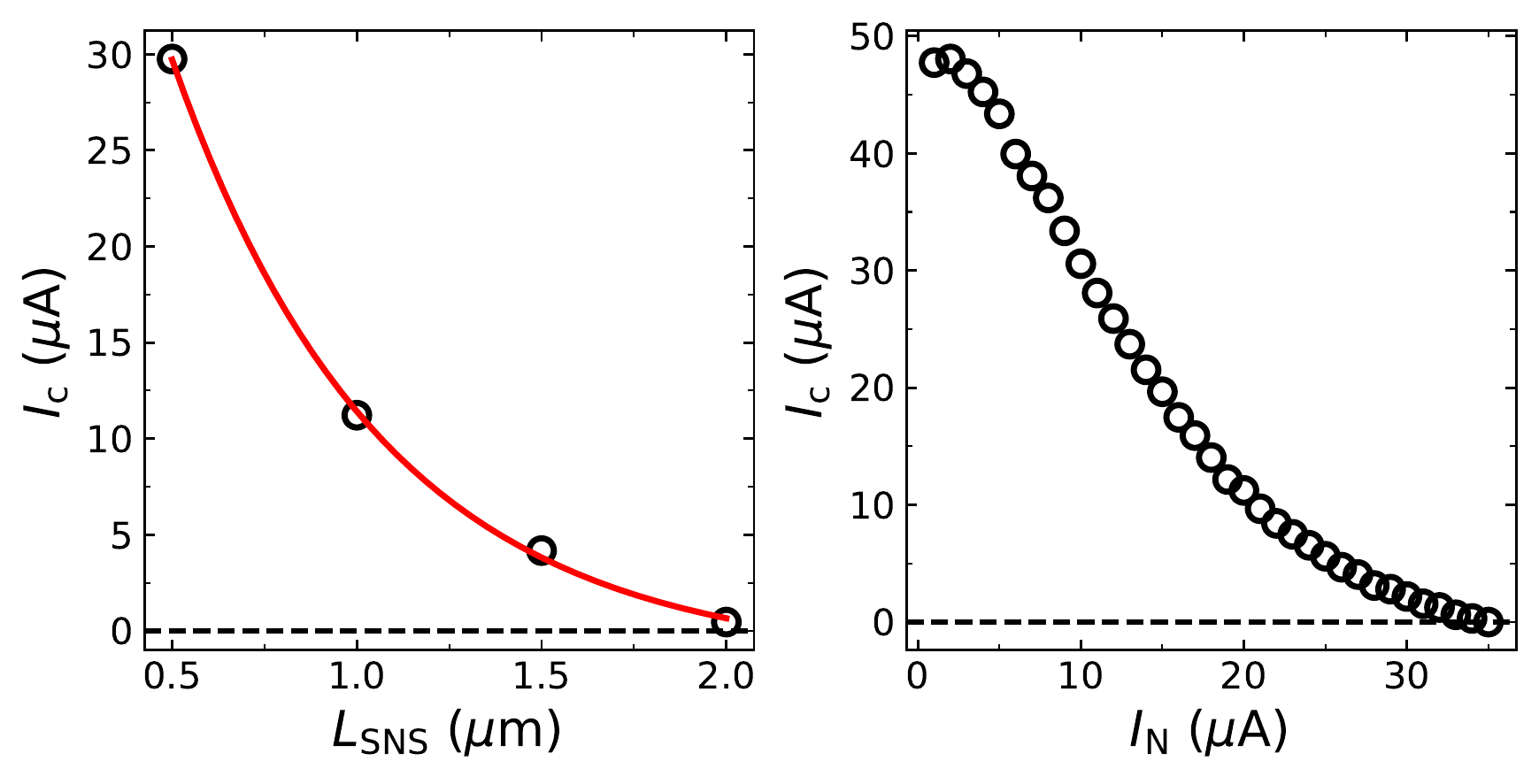}%
\caption{\label{fig:2}(a) Black open circles show the measured critical current as a function of $L_{\mathrm{SNS}}$ for SNS junctions with different $L_{\mathrm{SNS}}$. The critical current reduces as the length of the SNS junction is increased. The red line shows a fit to $I_{c}\propto\exp{(-L_{\mathrm{SNS}}/\xi)}$ from which an estimate of the coherence length can be made, $\xi\approx570\,\mathrm{nm}$. (b) Black open circles show the measured critical current of the SNS junction with increasing bias current $I_{\mathrm{N}}$ applied between the two normal reservoirs. The applied bias current $I_{\mathrm{N}}$ provides control of the critical current $I_{\mathrm{c}}$.} %
 \end{figure}
 Due to the proximity effect, a superconducting Josephson current can be induced in the normal segment of the HyQUID with a critical value that depends on the distance between the two superconducting contacts, $L_\mathrm{SNS}$. \cref{fig:2}(a) shows that as the distance between the superconducting contacts is increased the critical current of the SNS junction reduces. This allows specific critical current values to be designed during the fabrication stage.
It is also possible to modify the critical current of the SNS junction by applying a control current between the two normal reservoirs. \cref{fig:2}(b) shows the dependence of the critical current of an SNS junction on the applied bias current $I_{\mathrm{N}}$ allowing full suppression of $I_{\mathrm{c}}$ and thus control of the critical current during an experiment. For the HyQUID used in this work we design and fabricate the device parameters such that we are in the range $L_{\mathrm{SNS}}<\xi_{\mathrm{N}},L_{\phi}$. This allows us to modify the behaviour of the HyQUID during the experiment and access different operating regimes.

\subsection{HyQUID Fabrication and Measurement Setup}
Figures \ref{fig:1}(a) and \ref{fig:1}(b) show a scanning electron micrograph and circuit schematic of the interferometer described in this paper. The patterns were defined using standard electron beam lithography techniques. The normal and superconducting materials used were Ag and Al respectively. Both were deposited using thermal evaporation. In order to ensure a good interface between the two metals an \textit{in situ} argon etch was used. 

The SNS junction was 100 nm wide and 50 nm thick, and the distance between the two S contacts, $L_{\mathrm{SNS}} =500\,\mathrm{nm}$. The distance between the two N reservoirs, $L_{\mathrm{NNN}} =2000\,\mathrm{nm}$. Using a simple approximation for a square loop, we estimate the geometric inductance of the loop to be $L\approx100\,\mathrm{pH}$. All measurements were conducted in a $^{3}\mathrm{He}$ cryostat with a base temperature of 280\,mK. The resistance of the interferometer was measured using standard lock-in amplifier techniques. The magnetic flux through the interferometer $\Phi$ was controlled using both a superconducting solenoid in the cryostat and an on-chip flux line for pulsed measurements. The pulses were controlled using an arbitrary waveform generator. To modify the critical current we applied a dc current $I_{\mathrm{N}}$  between the two N reservoirs, so that $V_{\mathrm{N}} =I_{\mathrm{N}}  R_{\mathrm{N}}$.
We measure a normal state resistance $R_{\mathrm{N}}=2.8\,\Omega$ at 280\,mK. From this we calculate the diffusion coefficient $D=0.156\,\mathrm{m}^{2}\mathrm{s}^{-1}$. The normal state coherence length is then $\xi_{\mathrm{N}} = 780\,\mathrm{nm}$ and the phase breaking length $L_{\phi}=  6.93\,\mu\mathrm{m}$. Several interferometers were fabricated all showing similar behaviour.

\section{The HyQUID in Bifurcation Mode}
\subsection{HyQUID Dynamics}
\begin{figure}
\includegraphics[width=\columnwidth]{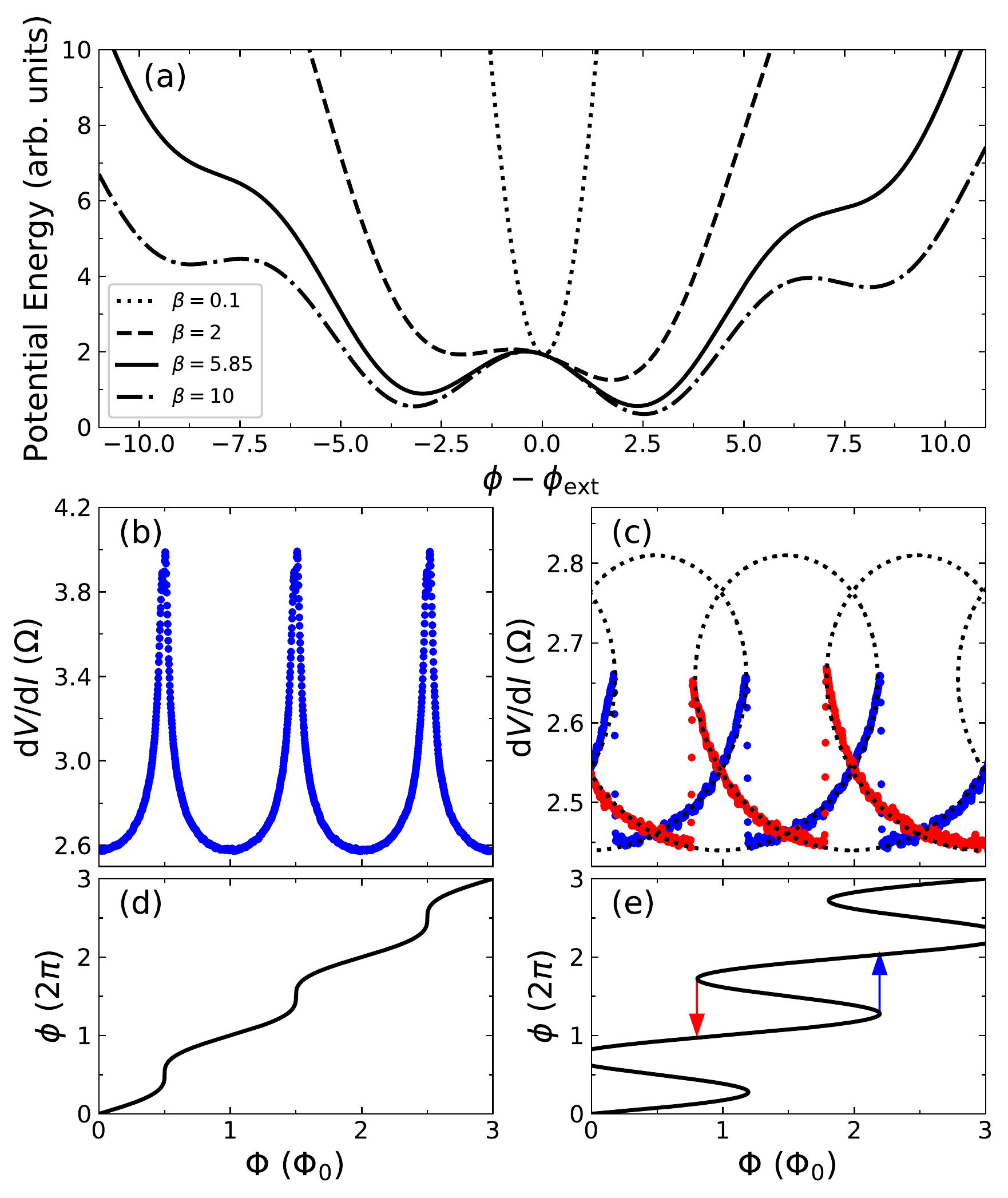}%
\caption{\label{fig:3}(a) The potential energy of the HyQUID (see \cref{eq:potential}) for different $\beta$ values. By varying $\beta$ the number and position of minima can be controlled. (b) and (c) show the differential resistance of the interferometer as a function of applied flux $\Phi$ for $I_{\mathrm{N}} =30\,\mu\mathrm{A}$ $(\beta=4.95)$ and $I_{\mathrm{N}} =50\,\mu\mathrm{A}$ $(\beta=1)$ respectively. The blue data shows the differential resistance when the flux is swept in the positive direction, red data when flux is swept in a negative direction. The black dotted line is a fit to theory using Eq.~\ref{eq:andreev} and \ref{eq:screening} showing the unstable regions of the curve. (d) and (e) show the phase across the interferometer $\phi$ as a function of the applied magnetic flux $\Phi$ calculated using the values of $\beta$ determined from (b) and (c) respectively. The blue and red arrows show the points where the phase switches between branches.}%
\end{figure}

Past investigations of hybrid normal-superconducting interferometers have focused on the range where $\xi_{\mathrm{N}}<L_{\mathrm{SNS}}<L_{\phi}$. In this regime the Josephson screening current in the flux sensitive loop of the interferometer is negligible. The phase-periodic oscillations in this regime are sinusoidal (or cusp-like) as shown in \cref{fig:3}(b), and the phase-flux relationship is single-valued as shown in 
\cref{fig:3}(d). 

In this work we investigate the HyQUID with $L_{\mathrm{SNS}}$ that is smaller than both the phase breaking length and the coherence length; $L_{\mathrm{SNS}}<\xi_{N},L_{\phi}$. In this regime the Josephson screening current is created with finite critical current $I_{c}$, changing the behaviour of the HyQUID drastically. The dynamics can be described using the resistively- and capacitively-shunted Josephson junction (RCSJ) model \cite{LikharevBook, clarke2006squid}. The dependence of the potential energy of our system is described by,
\begin{equation}
U=E_{\mathrm{J}}\left[1-\cos\phi+\frac{(\phi-\phi_{\mathrm{e}})^{2}}{2\beta} \right]
\label{eq:potential}
\end{equation}
where $\phi_{\mathrm{e}}$ is the externally applied phase, $E_{\mathrm{J}}=I_{\mathrm{c}}\Phi_{0}/2\pi$ is the Josephson energy, $\beta=2\pi LI_{\mathrm{c}}/\Phi_{0}$ is the screening parameter and $L$ is the geometric inductance of the loop.

The potential energy of the system can therefore be modified by control of $\beta$ (through varying the dc current $I_{\mathrm{N}}$) and through control of the external phase $\phi_{\mathrm{e}}$ applied to the HyQUID. 
At large values of $\beta$ (large critical currents) there are several stable states corresponding to minima in the local potential energy. At small values of $\beta$ (small critical currents) the system has only one stable minimum. \cref{fig:3}(a) shows the potential energy of the HyQUID for different $\beta$ values. The dynamics of such a system can be experimentally observed by measuring the differential resistance of the interferometer at different $\beta$ values. 

The Josephson screening current is created with finite critical current $I_{c}$, and the phase $\phi$ is related to the magnetic flux threading the loop by
\begin{equation}
2\pi\Phi/\Phi_{0}=\phi+\beta\sin\phi.
\label{eq:screening}
\end{equation}
As $\beta$ is increased the relationship between $\phi$ and $\Phi$ becomes increasingly non-linear and at $\beta>1$ becomes multi-valued. At the points shown by arrows in \cref{fig:3}(e) the system can exist in two different states (bifurcation). \cref{fig:3}(c) and (e) show the operating regime that is discussed in this paper. When $\beta>1$ the phase as a function of applied flux is multivalued - initially the superconducting phase starts on one branch with positive gradient. As the flux approaches a critical value the gradient changes sign. Beyond this critical flux the gradient becomes negative. The negative gradient represents values of phase that are unstable and thus the phase jumps to the next branch with a positive gradient. This results in dynamics that are path dependent (in our case, flux sweep direction dependent). The differential resistance as a function of applied flux is shown in \cref{fig:3}(c). The blue data is the positive flux sweep, the red data is the negative flux sweep. The blue and red arrows on the phase-flux diagram of \cref{fig:3}(e) represent the corresponding points where the phase switches between branches. It should be noted that the two regimes shown (\cref{fig:3}(b) and \cref{fig:3}(d)) are measured using the same HyQUID. Only the applied dc current $I_{\mathrm{N}}$ is used to tune between these regimes. The full range of HyQUID differential resistance behaviour is shown in \cref{app:tuning}.

\begin{figure}[h!]
\includegraphics[width=\columnwidth]{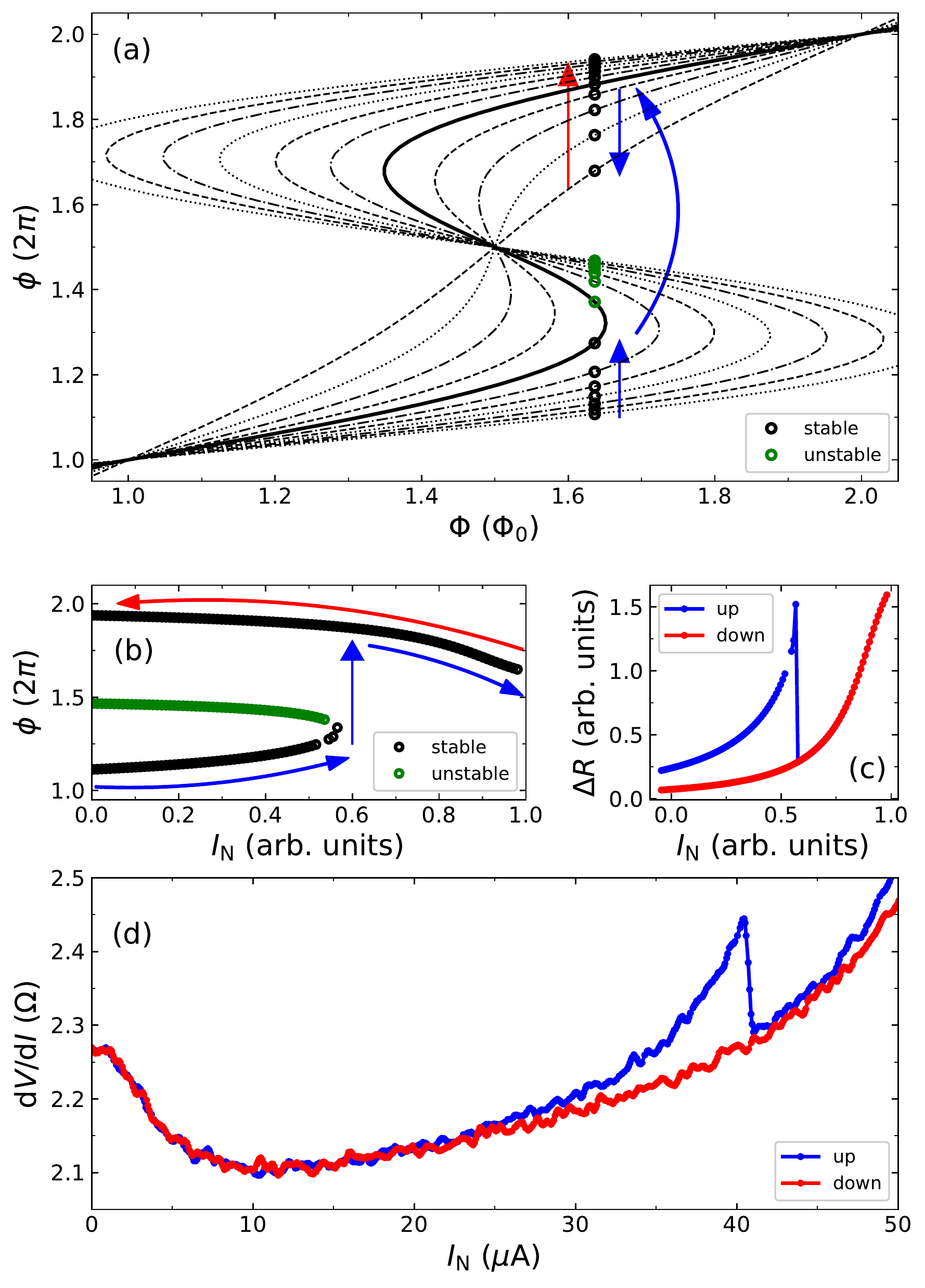}%
\caption{\label{fig:4}(a) Phase-flux diagram for multiple values of $\beta$. For clarity only a subset of curves are shown. The external flux is fixed ($\Phi = 1.64\Phi_{0}$) and the phase $\phi$ is found for each $\beta$ by solving \cref{eq:screening}. Some of the phase solutions occur on negative gradients (green open circles) and are unstable. Only the stable solutions (black open circles) are experimentally accessible. The path dependence of the dynamics is illustrated on the graph; as the bias current is increased ($\beta$ decreased) the phase increases until a negative gradient of the phase-flux curve is reached (occurs on the solid black curve). The phase then jumps to the next stable level and then progresses smoothly again. This can be observed by following the blue arrows on the graph. The red arrow represents the phase progression as the bias current is reduced ($\beta$ increased). (b) The phase as a function of the applied dc current. As the current is increased a jump in the phase is observed. (c) This phase jump is then observed in the calculated differential resistance of the HyQUID (\cref{eq:andreev}). A clear path dependence is observed based on the direction of bias current sweep. (d) Experimental differential resistance data as a function of the applied dc bias current. Blue is the \textit{up} current sweep, red is the \textit{down} current sweep. The behaviour of the HyQUID during the experiment follows the model presented in (a)-(c).}%
\end{figure}

The differential resistance close to the bistable points shows an extreme sensitivity to applied flux (as per \cref{fig:3}(c)). This happens because the height of the potential energy barrier keeping the system in the local potential well becomes small enough so that the system escapes the well. As we see from \cref{eq:potential} the escape can be stimulated by manipulation of external flux or, alternatively, by manipulation of $\beta$ via applied dc current $I_{\mathrm{N}}$ (as per \cref{fig:3}(a)). The differential resistance as a function of the bias current $I_{\mathrm{N}}$ is shown in \cref{fig:4}. In this control protocol the applied flux to the HyQUID is fixed and the bias current $I_{\mathrm{N}}$ is swept. Similar to the protocol of \cref{fig:3} the dynamics arise as a result of instabilities of the phase-flux diagram. In this case the value of $\beta$ varies as  $I_{\mathrm{N}}$ is varied, therefore the form of the phase-flux dependence varies as $I_{\mathrm{N}}$ is varied.

As $I_{\mathrm{N}}$ is increased $\beta$ reduces. As $\beta$ is reduced the phase-flux diagram tends towards linear. For a fixed applied flux as shown in \cref{fig:4}(a), the HyQUID phase increases as $I_{\mathrm{N}}$ is increased. At some value of $I_{\mathrm{N}}$ the phase becomes unstable and a phase jump to the next branch occurs. The phase progression as $I_{\mathrm{N}}$ is varied is shown in \cref{fig:4}(a-b). The blue arrows represent the path when $I_{\mathrm{N}}$ is increased. The red arrows represent the path when  $I_{\mathrm{N}}$ is decreased. By considering the differential resistance of the HyQUID in this regime we again find that a hysteretic behaviour is observed (\cref{fig:4}(c)) which allows the HyQUID to be used as a latched readout device. The experimental dependence of the differential resistance as a function of  $I_{\mathrm{N}}$ for our device is shown in \cref{fig:4}(d) and agrees qualitatively with the modelled data.

\subsection{Latching Readout}
As described in the previous sections the high sensitivity of the HyQUID to small changes in flux and/or bias current at the bifurcation points can be used for the construction of a new type of latching amplifier that can be used for the readout of quantum circuits, such as superconducting flux qubits.

To use the HyQUID as a latching readout we take advantage of the bistability in the HyQUID dynamics and implement pulsed measurements using two alternative protocols for manipulation of flux and bias current.
\subsubsection{Latching readout using flux manipulation}

\begin{figure}
 \includegraphics[width=\columnwidth]{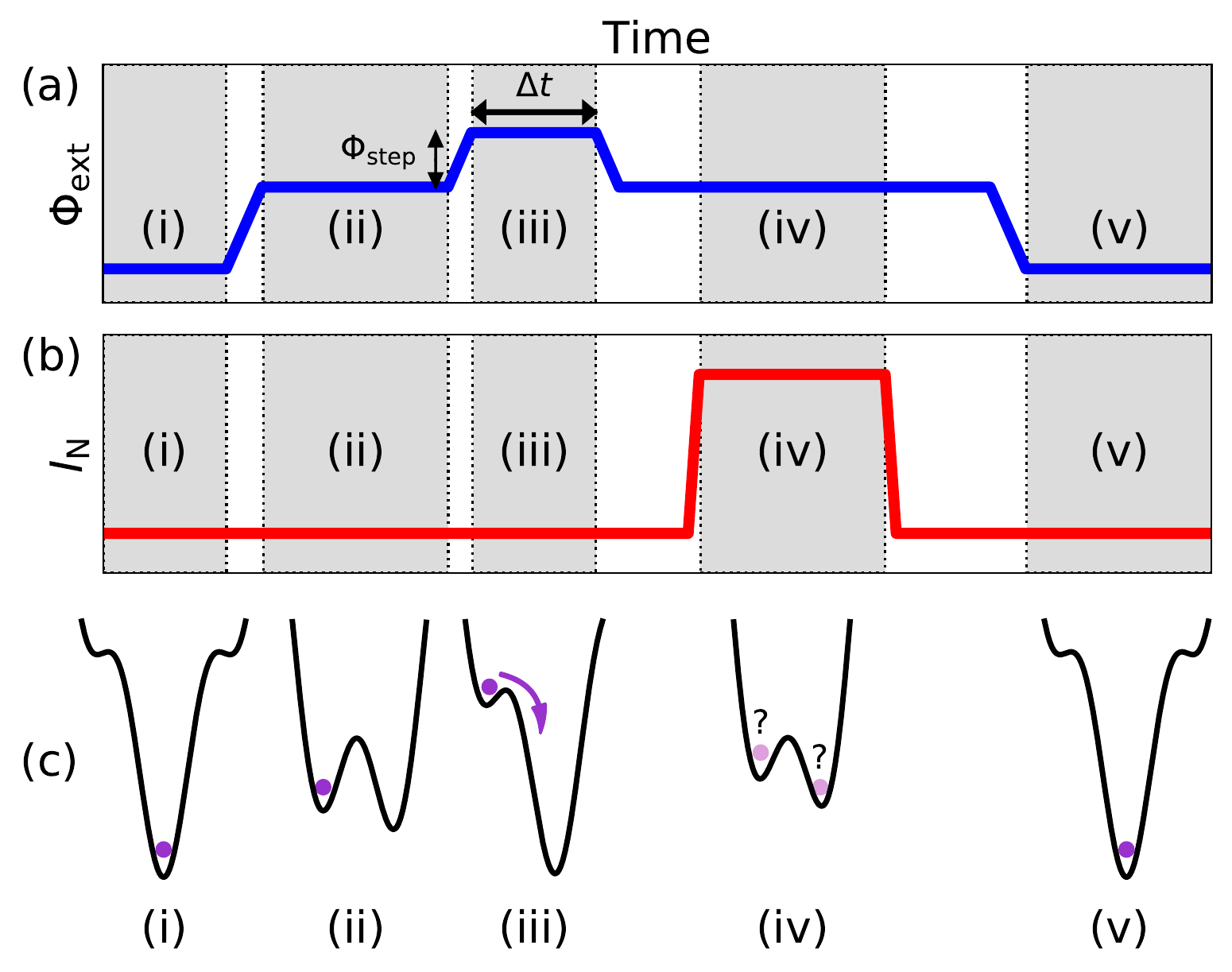}%
 \caption{\label{fig:5}Description of the pulsed measurement protocol used to measure escape rates from the systems potential well. (a) The blue line shows the applied flux $\Phi{\mathrm{ext}}$ as a function of time. (b) The red line shows the measurement current $I_{\mathrm{N}}$ through the vertical branch of the cross. The protocol can be described by five regions (i) through (v) with the energy potential shown for each region in (c): (i) System is initialized so that it sits in a lower well. (ii) The well is raised by increasing $\Phi_{\mathrm{ext}}$. (iii) A short flux pulse $\Phi_{\mathrm{step}}$ is applied for time $\Delta t$, to temporarily lower the barrier. (iv) The barrier is reinstated by reducing $\Phi_{\mathrm{ext}}$. The HyQUID resistance is measured to determine if the system is in the upper or lower well. (v) The system is reset. The escape probability $P_{\mathrm{escape}}$ can be determined by repeating the process as a function of $\Phi_{\mathrm{step}}$.}%
 \end{figure}

The readout protocol is shown in \cref{fig:5}. The applied flux $\Phi_{\mathrm{ext}}$ to the HyQUID is controlled via an on-chip flux line. The dc current $I_{\mathrm{N}}$ is added to the small ac-current used for lock-in detection. Both pulses are controlled using an arbitrary waveform generator. We again use the RCSJ model description of our system and use the pulsed protocol to control the `particle' in a potential energy well. The system is initialised so that the particle sits in a single lower well (\cref{fig:5}(c-i)). By increasing the applied flux the well is raised (\cref{fig:5}(c-ii)) trapping the particle. A short flux pulse $\Phi_{\mathrm{step}}$ is applied for time $\Delta t$, temporarily lowering the barrier between the two wells (\cref{fig:5}(c-iii)) allowing for the possibility that the particle escapes to the other well - this leads to a sudden change in phase (which manifests as a change to the HyQUID resistance). The barrier is reinstated by reducing the applied flux and a measurement of the HyQUID resistance is made - due to the latching nature of operation this measurement determines whether or not the particle escaped (\cref{fig:5}(c-iv)). Finally, the system is reset by removing the flux and current bias (\cref{fig:5}(c-v)).

During stage (iii) of the measurement protocol the barrier is reduced and the particle has some possibility of escaping the the lower well. The escape rate from the potential well is written as 
\begin{equation}
\Gamma=\frac{\omega}{2\pi}\exp[-U_{0}/k_{\mathrm{B}}T]
\end{equation}
where $\omega$ is the escape attempt frequency and $U_{0}$ is the barrier height \cite{Kofman_PRB_2007}. For shallow wells, $U_{0}$ can be described by the cubic approximation
\begin{equation}
\label{eq:potentialU0}
U_{0}=\frac{2}{3}\sqrt{1-\beta^{-2}}E_{\mathrm{J}}\epsilon^{3}
\end{equation}
where
\begin{equation}
\epsilon=\sqrt{2(\phi_{\mathrm{c}}-\phi)/\sqrt{\beta^{2}-1}}
\end{equation}
and $\phi_{\mathrm{c}}=[\pi/2+\sqrt{\beta^{2}-1}+\sin^{-1}(1/\beta)]$ is the external phase at which $U_{0}=0$. 
Since we ensure that the particle always starts in the upper well we can write $P_{\mathrm{escape}} =1-e^{-\Gamma\Delta t}$.

\begin{figure}[!t]
\includegraphics[width=\columnwidth]{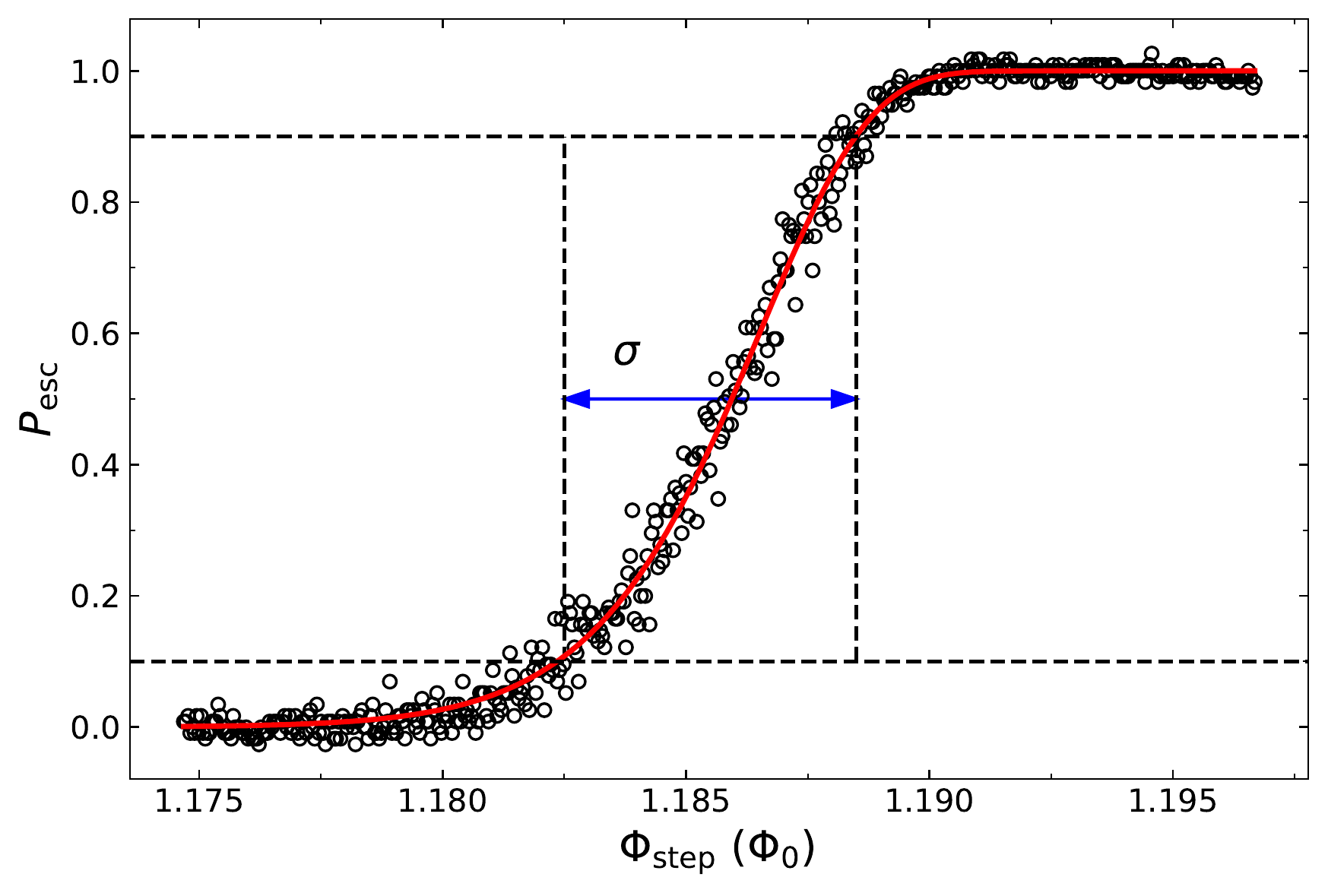}
\caption{\label{fig:6} Probability of escape as a function of step height measured using the method described in \cref{fig:5} using a pulse length $\Delta t=200\,\mu\mathrm{s}$. $\sigma$ is a figure of merit describing the difference in $\Phi_{\mathrm{step}}$  between $P_{\mathrm{escape}}=0.1$ and $P_{\mathrm{escape}}=0.9$. For this system $\sigma=0.006\,\Phi_{0}$.
 }%
\end{figure}

By repeating the measurement protocol many times and stepping the height of the applied flux pulse that controls the barrier a probability curve of state occupation can be experimentally determined. The resulting probability curve is shown in \cref{fig:6}. We define $\sigma$ as a figure of merit describing the fidelity of the readout, which is the difference in flux $\Phi$ between $P_{\mathrm{escape}}=0.1$ and $P_{\mathrm{escape}}=0.9$. We show that our HyQUID readout implementation can detect changes in flux $\sigma = 0.006\,\Phi_{0}$. This resolution is comparable to that of the bifurcation amplifier used by Lupa{\c s}cu \textit{et al}.~to probe the state of a flux qubit \cite{Lupascu_PRL_2006}. 

\subsubsection{Latching readout using bias current manipulation}

\begin{figure}
 \includegraphics[width=\columnwidth]{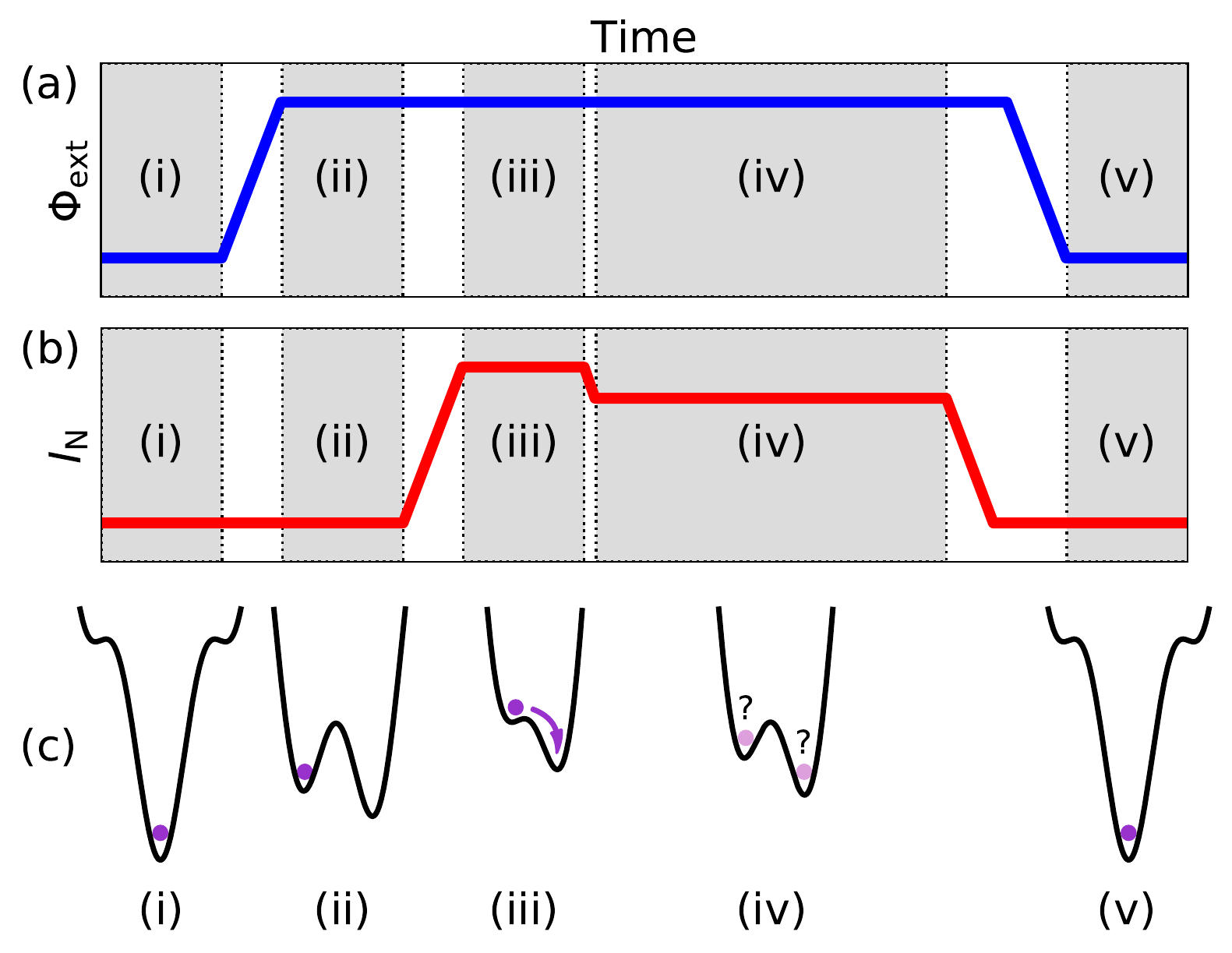}%
 \caption{\label{fig:7} Alternative pulsed measurement protocol to determine the escape probability. (a) Applied flux $\Phi_{\mathrm{ext}}$ as a function of time. (b) Current $I_{\mathrm{N}}$ as a function of time. Similar to \cref{fig:5} the protocol can be described by five regions (i) through (v) with the energy potential shown for each region in (c): (i) System is initialized so that it sits in a lower well. (ii) The well is raised by increasing $\Phi_{\mathrm{ext}}$. (iii) A short measurement current pulse is applied to temporarily lower the barrier. (iv) The barrier is reinstated by reducing the measurement current pulse. The HyQUID resistance is measured to determine if the system is in the upper or lower well. (v) The system is reset. The escape probability $P_{\mathrm{escape}}$ can be determined by repeating the process as a function of $I_{\mathrm{N}}$.}%
 \end{figure}

\begin{figure}
\includegraphics[width=\columnwidth]{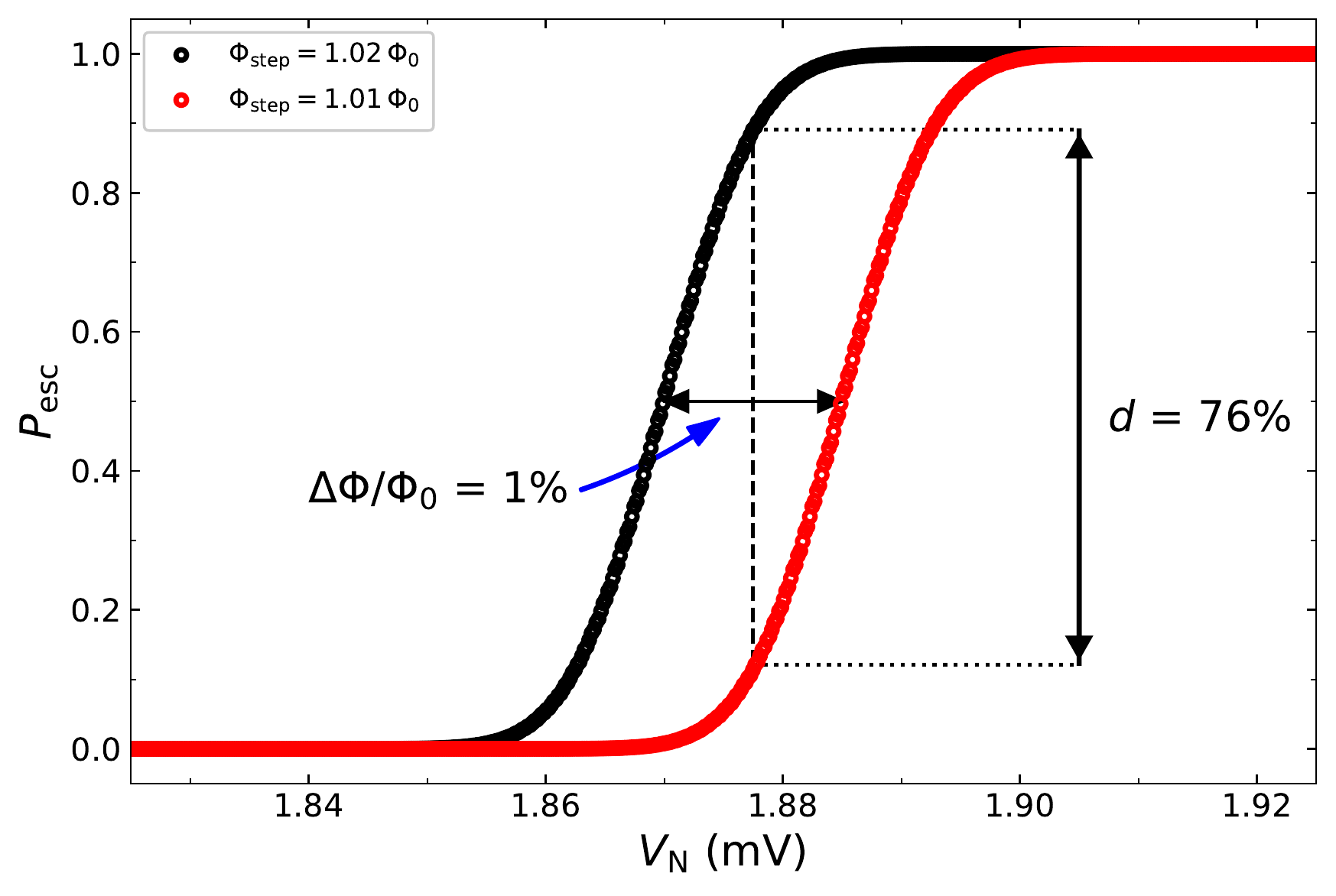}%
 \caption{\label{fig:8} Probability of escape as a function of voltage applied across the normal reservoirs. We show $P_{\mathrm{esc}}$ vs $V_{\mathrm{N}}$ curves corresponding to two flux positions. The curves are obtained using the experimental $P_{\mathrm{esc}}$ vs $\Phi_{\mathrm{step}}$ curves like shown in \cref{fig:6}. We follow the convention of Siddiqi \textit{et al.}~\cite{Siddiqi_PRL_2004} and describe the discrimination power $d$ as the maximum difference between two switching probability curves. For $\Delta\Phi/\Phi_{0}=1\%$ we find $d\approx76\%$.}%
 \end{figure}

The protocol shown in \cref{fig:7} uses the bias current $I_{\mathrm{N}}$ to control the potential well (by controlling $\beta$). One advantage of this protocol is that the applied flux is constant after initialisation which could be beneficial when the readout circuit is coupled to a flux-sensitive device such as a flux qubit. To estimate the performance of the HyQUID latching readout using the protocol shown in \cref{fig:7} we use experimentally determined probability curves for multiple fixed values of $I_{\mathrm{N}}$. From the experimental data shown in \cref{fig:6} we fit the probability curve and extrapolate between probability curves measured at different $I_{\mathrm{N}}$. This allows the simulation of a probability curve for the protocol shown in \cref{fig:7}. The switching probabilities at two different values of applied flux are shown in \cref{fig:8}. We show two flux values $\Phi_{\mathrm{step}} = 1.02\,\Phi_{0}$, and $\Phi_{\mathrm{step}} = 1.01\,\Phi_{0}$. Following the convention from Siddiqi \textit{et al.}~\cite{Siddiqi_PRL_2004} we define the discrimination power $d$ as the maximum difference between the two switching probability curves. The relative difference between the two curves can be described as $\Delta\Phi = 2(\Phi_{2}-\Phi_{1}) /(\Phi_{2}+\Phi_{1})$. We find that for $\Delta\Phi/\Phi_{0}\approx1\%$, $d=76\%$. This is comparable to the performance described in Ref \onlinecite{Siddiqi_PRL_2004}.
\section{Discussion and Conclusion}

The latching amplifier is suitable for probing the state of a superconducting qubit for a number of reasons. Firstly, by correctly tuning the value of $\beta$ it is possible to make the magnetometer sensitive to extremely small changes in flux. Secondly due to its latching action, the act of reading out the qubit can be separated from the act of probing the qubit, minimising the back-action. Thirdly, by minimising the mutual inductance between the flux sensitive loop and the bias current/voltage leads (for instance, using a HyQUID with a folded cross design as per Ref \onlinecite{Shelly2016}) the decoherence of the qubit by the measurement system can be minimised. 

An important advantage of the HyQUID in bifurcation mode is that the screening Josephson current acts to reduce the thermal flux noise $\delta \Phi_{\mathrm{th}}$ introduced by the Nyquist noise in the SNS junction. In the absence of Josephson screening current the Nyquist thermal noise current $\delta I_{\mathrm{th}}=\sqrt{4k_{\mathrm{B}}T\Delta f/R}$, where $R$ is the resistance of the SNS junction in the normal state, introduces a current that circulates in the loop $\delta I_{i} = \delta I_{\mathrm{th}}$. The thermal flux noise in the loop is then given by
\begin{equation}
\delta \Phi_{\mathrm{th}} = L\delta I_{\mathrm{th}}.
\end{equation}
This thermal flux noise increases the noise floor which reduces the sensitivity of the magnetometer and contributes to the back-action of the readout.

The Josephson supercurrent $\delta I_{s}$ influences the current $\delta I_{i}$ and partially screens the Nyquist current. The current circulating in the loop is $\delta I_{i} = \delta I_{s}+\delta I_{\mathrm{th}}$ due to current conservation \cite{Shelly2016}. The value of $\delta I_{i}$ can be determined by minimising the energy of the loop \cite{Gurevich2006,Shelly2016}. The total energy of the loop is,
\begin{equation}
W = W_{k} + W_{m} = \frac{L\delta I_{i}^{2}}{2} + \frac{L_{k}\delta I_{s}^{2}}{2} 
\end{equation}
where the first term describes the energy of the magnetic field due to the circulating current $\delta I_{i}$. The second term describes the kinetic energy due to the superconducting electrons in the normal branch of the HyQUID.
The resulting thermal flux noise is given by,
\begin{equation}
\delta \Phi_{\mathrm{th}} = \left(\frac{LL_{k}}{L+L_{k}}\right)I_{\mathrm{th}},
\end{equation}
where $L_{k}=\Phi_{0}/2\pi I_{c}$ is the kinetic inductance. Note that $\delta \Phi_{\mathrm{th}}\rightarrow L_{k}I_{\mathrm{th}}\rightarrow0$ as $I_{c}\rightarrow\infty$ therefore designing a HyQUID with a high critical current acts to reduces the flux noise induced by the interferometer that can cause decoherence in the coupled qubit.

To further optimize the HyQUID for latching readout we focus on the figures of merit $\sigma$ and $d$. By examination of \cref{eq:potentialU0} we see that the value of $\mathrm{d}U_{0}/\mathrm{d}\Phi$ close to $\Phi=\Phi_{0}/2\pi$ should reach a maximum as $\beta\rightarrow 1$. Therefore to minimise $\sigma$, the HyQUID should be operated close to this point. The performance of the HyQUID could therefore be improved by designing an interferometer with  $1>\beta>1.5$ when $I_{\mathrm{N}}=0$. This could be achieved by adjusting the junction length $L_{\mathrm{SNS}}$  and/or the HyQUID loop size. This would allow the HyQUID to operate at close to its optimal point using only a small bias current $I_{\mathrm{N}}$ through the resistive N branch reducing any heating effects. The response time of the HyQUID is estimated to be less than 40 ps \cite{Checkley_2011} - fast enough to enable utilization of the HyQUID latching readout in typical qubit readout protocols (see for instance \cite{Barends2014,Arute2019}).

In conclusion, we present a Hybrid Quantum Interference Device that can be set in a bistable regime through \textit{in situ} control of the Josephson screening current by application of a dc tuning current without entering the normal state. We show that in this regime the HyQUID can be operated in a latching mode for quantum circuit readout. We have employed a pulsed measurement to investigate the dynamics of the HyQUID in this mode, and test the fidelity of the readout. The differential resistance behaviour as a function of either flux or current is in agreement with our modelling of the HyQUID dynamics. 

We believe that this embodiment of the HyQUID is suitable for applications where high sensitivity and low back-action are needed. In particular, we believe that the techniques described here could be applied to the readout of superconducting flux qubits.

The latching dynamics of the HyQUID may also have utility in superconducting logic architectures as a storage cell such as those proposed in Ref \onlinecite{ligato2020persistent} where a quantum state may stored and then read out much later.

\acknowledgments
CDS thanks J.~J.~Burnett for useful discussion related to qubit readout strategies, and P.~J.~Meeson for HyQUID discussion. This work was supported by the Engineering and Physical Sciences Research Council (UK) Grant EP/E012469/1. CDS gratefully acknowledges the UK Department of Business, Energy and Industrial Strategy (BEIS) for funding during the writing of this manuscript.

\appendix

\section{Tunable HyQUID Operation}
\label{app:tuning}

\begin{figure}[h!]
\includegraphics[width=\columnwidth]{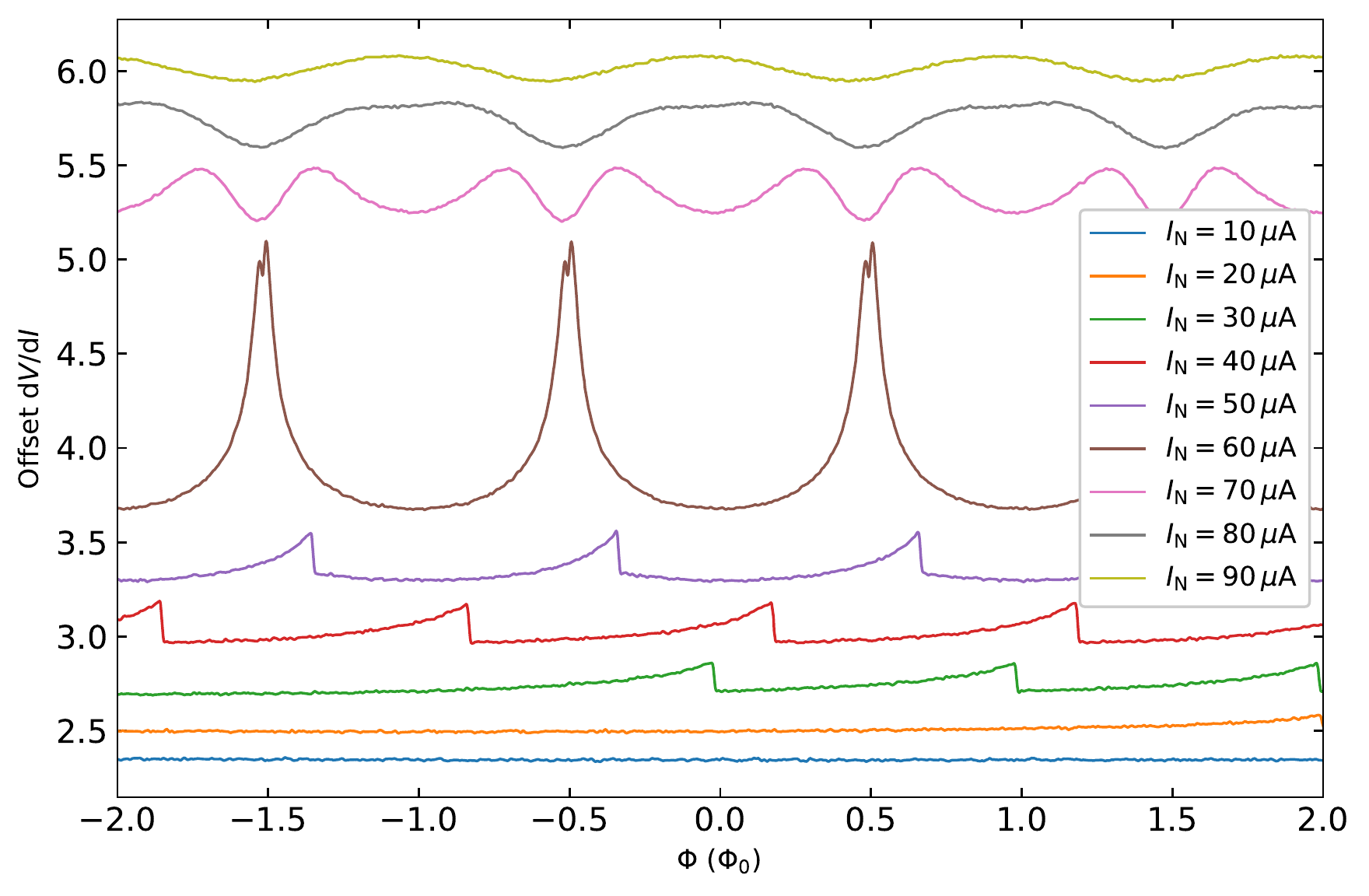}
\caption{\label{fig:app1}(a) Differential resistance of the HyQUID as a function of applied flux  $\Phi$ at different bias currents $I_{N}$. The plots are offset on the y-axis for clarity. Progression from the hysteretic regime ($I_{\mathrm{N}}$ large; $I_{\mathrm{c}},\,\beta$ small) to the non-hysteretic regime ($I_{\mathrm{N}}$ small; $I_{\mathrm{c}},\,\beta$ large) is clearly demonstrated.}%
\end{figure}

As discussed in the main text the HyQUID operation regime can be tuned by varying the $\beta$ screening parameter (by varying the critical current through application of a dc current between the two normal reservoirs). Figure \ref{fig:app1} shows the oscillations of the differential resistance of the interferometer as a function of the applied flux $\Phi$ at different  $I_{\mathrm{N}}$-controlled $\beta$. At large $\beta$ periodic oscillations with regular sharp changes in the resistance as a function of the applied flux are seen showing hysteresis when the field sweep direction is reversed. The phase as a function of flux at this value of $\beta$ is multivalued. Initially the superconducting phase starts on one branch where there is a relatively small gradient. As the flux approaches a critical value the gradient changes sign. Past this critical flux, the gradient becomes negative. Values of phase on the negative gradient are unstable and so the phase jumps to the next branch, leading to the sudden switch seen in the magnetoresistance. This process is described in the main text and shown in \cref{fig:3}. 

As we increase $I_{\mathrm{N}}$, the hysteresis disappears (corresponds to the interferometer having  $\beta < 1$). The relationship between phase and flux is no longer multivalued and $R_{\mathrm{N}}$ as a function of $\Phi$ is no longer path dependant (see Figures \ref{fig:3}(b) and \ref{fig:3}(d)). The transfer function $\mathrm{d}V/\mathrm{d}\Phi$ is maximised around $\Phi=\Phi_{0}/2$. Operating the interferometer in this regime enhances the sensitivity with $\mathrm{d}V/\mathrm{d}\Phi=1.92\,\mu\mathrm{V}\,\Phi_{0}^{-1}$ (where $V$ is the voltage measured by the lock-in amplifier across $R_{\mathrm{N}}$) compared to $0.27\,\mu\mathrm{V},\Phi_{0}^{-1}$ for a perfectly sinusoidal oscillation of the same amplitude.

At higher values of $I_{\mathrm{N}}$  we observe a splitting of the peaks of the magnetoresistance oscillations and eventually a $\pi$ shift in the phase of the oscillations. This is a consequence of the change in the differential resistance at higher measuring currents. A detailed explanation of this phenomenon can be found in Ref \cite{Petrashov_PRB_1998}.

\bibliography{hyst2020}

\begin{thebibliography}{23}%
\makeatletter
\providecommand \@ifxundefined [1]{%
 \@ifx{#1\undefined}
}%
\providecommand \@ifnum [1]{%
 \ifnum #1\expandafter \@firstoftwo
 \else \expandafter \@secondoftwo
 \fi
}%
\providecommand \@ifx [1]{%
 \ifx #1\expandafter \@firstoftwo
 \else \expandafter \@secondoftwo
 \fi
}%
\providecommand \natexlab [1]{#1}%
\providecommand \enquote  [1]{``#1''}%
\providecommand \bibnamefont  [1]{#1}%
\providecommand \bibfnamefont [1]{#1}%
\providecommand \citenamefont [1]{#1}%
\providecommand \href@noop [0]{\@secondoftwo}%
\providecommand \href [0]{\begingroup \@sanitize@url \@href}%
\providecommand \@href[1]{\@@startlink{#1}\@@href}%
\providecommand \@@href[1]{\endgroup#1\@@endlink}%
\providecommand \@sanitize@url [0]{\catcode `\\12\catcode `\$12\catcode
  `\&12\catcode `\#12\catcode `\^12\catcode `\_12\catcode `\%12\relax}%
\providecommand \@@startlink[1]{}%
\providecommand \@@endlink[0]{}%
\providecommand \url  [0]{\begingroup\@sanitize@url \@url }%
\providecommand \@url [1]{\endgroup\@href {#1}{\urlprefix }}%
\providecommand \urlprefix  [0]{URL }%
\providecommand \Eprint [0]{\href }%
\providecommand \doibase [0]{http://dx.doi.org/}%
\providecommand \selectlanguage [0]{\@gobble}%
\providecommand \bibinfo  [0]{\@secondoftwo}%
\providecommand \bibfield  [0]{\@secondoftwo}%
\providecommand \translation [1]{[#1]}%
\providecommand \BibitemOpen [0]{}%
\providecommand \bibitemStop [0]{}%
\providecommand \bibitemNoStop [0]{.\EOS\space}%
\providecommand \EOS [0]{\spacefactor3000\relax}%
\providecommand \BibitemShut  [1]{\csname bibitem#1\endcsname}%
\let\auto@bib@innerbib\@empty
\bibitem [{\citenamefont {Chiorescu}\ \emph {et~al.}(2003)\citenamefont
  {Chiorescu}, \citenamefont {Nakamura}, \citenamefont {Harmans},\ and\
  \citenamefont {Mooij}}]{Chiorescu_Science_2003}%
  \BibitemOpen
  \bibfield  {author} {\bibinfo {author} {\bibfnamefont {I.}~\bibnamefont
  {Chiorescu}}, \bibinfo {author} {\bibfnamefont {Y.}~\bibnamefont {Nakamura}},
  \bibinfo {author} {\bibfnamefont {C.~J. P.~M.}\ \bibnamefont {Harmans}}, \
  and\ \bibinfo {author} {\bibfnamefont {J.~E.}\ \bibnamefont {Mooij}},\ }\href
  {\doibase 10.1126/science.1081045} {\bibfield  {journal} {\bibinfo  {journal}
  {Science}\ }\textbf {\bibinfo {volume} {299}},\ \bibinfo {pages} {1869}
  (\bibinfo {year} {2003})}\BibitemShut {NoStop}%
\bibitem [{\citenamefont {Martinis}\ \emph {et~al.}(2009)\citenamefont
  {Martinis}, \citenamefont {Ansmann},\ and\ \citenamefont
  {Aumentado}}]{Martinis_PRL_2009}%
  \BibitemOpen
  \bibfield  {author} {\bibinfo {author} {\bibfnamefont {J.~M.}\ \bibnamefont
  {Martinis}}, \bibinfo {author} {\bibfnamefont {M.}~\bibnamefont {Ansmann}}, \
  and\ \bibinfo {author} {\bibfnamefont {J.}~\bibnamefont {Aumentado}},\ }\href
  {\doibase 10.1103/PhysRevLett.103.097002} {\bibfield  {journal} {\bibinfo
  {journal} {Phys. Rev. Lett.}\ }\textbf {\bibinfo {volume} {103}},\ \bibinfo
  {pages} {097002} (\bibinfo {year} {2009})}\BibitemShut {NoStop}%
\bibitem [{\citenamefont {M\"annik}\ and\ \citenamefont
  {Lukens}(2004)}]{Mannik_PRL_2004}%
  \BibitemOpen
  \bibfield  {author} {\bibinfo {author} {\bibfnamefont {J.}~\bibnamefont
  {M\"annik}}\ and\ \bibinfo {author} {\bibfnamefont {J.~E.}\ \bibnamefont
  {Lukens}},\ }\href {\doibase 10.1103/PhysRevLett.92.057004} {\bibfield
  {journal} {\bibinfo  {journal} {Phys. Rev. Lett.}\ }\textbf {\bibinfo
  {volume} {92}},\ \bibinfo {pages} {057004} (\bibinfo {year}
  {2004})}\BibitemShut {NoStop}%
\bibitem [{\citenamefont {Lupascu}\ \emph {et~al.}(2006)\citenamefont
  {Lupascu}, \citenamefont {Driessen}, \citenamefont {Roschier}, \citenamefont
  {Harmans},\ and\ \citenamefont {Mooij}}]{Lupascu_PRL_2006}%
  \BibitemOpen
  \bibfield  {author} {\bibinfo {author} {\bibfnamefont {A.}~\bibnamefont
  {Lupascu}}, \bibinfo {author} {\bibfnamefont {E.~F.~C.}\ \bibnamefont
  {Driessen}}, \bibinfo {author} {\bibfnamefont {L.}~\bibnamefont {Roschier}},
  \bibinfo {author} {\bibfnamefont {C.~J. P.~M.}\ \bibnamefont {Harmans}}, \
  and\ \bibinfo {author} {\bibfnamefont {J.~E.}\ \bibnamefont {Mooij}},\ }\href
  {\doibase 10.1103/PhysRevLett.96.127003} {\bibfield  {journal} {\bibinfo
  {journal} {Phys. Rev. Lett.}\ }\textbf {\bibinfo {volume} {96}},\ \bibinfo
  {pages} {127003} (\bibinfo {year} {2006})}\BibitemShut {NoStop}%
\bibitem [{\citenamefont {Siddiqi}\ \emph {et~al.}(2004)\citenamefont
  {Siddiqi}, \citenamefont {Vijay}, \citenamefont {Pierre}, \citenamefont
  {Wilson}, \citenamefont {Metcalfe}, \citenamefont {Rigetti}, \citenamefont
  {Frunzio},\ and\ \citenamefont {Devoret}}]{Siddiqi_PRL_2004}%
  \BibitemOpen
  \bibfield  {author} {\bibinfo {author} {\bibfnamefont {I.}~\bibnamefont
  {Siddiqi}}, \bibinfo {author} {\bibfnamefont {R.}~\bibnamefont {Vijay}},
  \bibinfo {author} {\bibfnamefont {F.}~\bibnamefont {Pierre}}, \bibinfo
  {author} {\bibfnamefont {C.~M.}\ \bibnamefont {Wilson}}, \bibinfo {author}
  {\bibfnamefont {M.}~\bibnamefont {Metcalfe}}, \bibinfo {author}
  {\bibfnamefont {C.}~\bibnamefont {Rigetti}}, \bibinfo {author} {\bibfnamefont
  {L.}~\bibnamefont {Frunzio}}, \ and\ \bibinfo {author} {\bibfnamefont
  {M.~H.}\ \bibnamefont {Devoret}},\ }\href {\doibase
  10.1103/PhysRevLett.93.207002} {\bibfield  {journal} {\bibinfo  {journal}
  {Phys. Rev. Lett.}\ }\textbf {\bibinfo {volume} {93}},\ \bibinfo {pages}
  {207002} (\bibinfo {year} {2004})}\BibitemShut {NoStop}%
\bibitem [{\citenamefont {Siddiqi}\ \emph {et~al.}(2006)\citenamefont
  {Siddiqi}, \citenamefont {Vijay}, \citenamefont {Metcalfe}, \citenamefont
  {Boaknin}, \citenamefont {Frunzio}, \citenamefont {Schoelkopf},\ and\
  \citenamefont {Devoret}}]{Siddiqi_PRB_2006}%
  \BibitemOpen
  \bibfield  {author} {\bibinfo {author} {\bibfnamefont {I.}~\bibnamefont
  {Siddiqi}}, \bibinfo {author} {\bibfnamefont {R.}~\bibnamefont {Vijay}},
  \bibinfo {author} {\bibfnamefont {M.}~\bibnamefont {Metcalfe}}, \bibinfo
  {author} {\bibfnamefont {E.}~\bibnamefont {Boaknin}}, \bibinfo {author}
  {\bibfnamefont {L.}~\bibnamefont {Frunzio}}, \bibinfo {author} {\bibfnamefont
  {R.~J.}\ \bibnamefont {Schoelkopf}}, \ and\ \bibinfo {author} {\bibfnamefont
  {M.~H.}\ \bibnamefont {Devoret}},\ }\href {\doibase
  10.1103/PhysRevB.73.054510} {\bibfield  {journal} {\bibinfo  {journal} {Phys.
  Rev. B}\ }\textbf {\bibinfo {volume} {73}},\ \bibinfo {pages} {054510}
  (\bibinfo {year} {2006})}\BibitemShut {NoStop}%
\bibitem [{\citenamefont {Vijay}\ \emph {et~al.}(2009)\citenamefont {Vijay},
  \citenamefont {Devoret},\ and\ \citenamefont
  {Siddiqi}}]{Vijay_RevSciInst_2009}%
  \BibitemOpen
  \bibfield  {author} {\bibinfo {author} {\bibfnamefont {R.}~\bibnamefont
  {Vijay}}, \bibinfo {author} {\bibfnamefont {M.~H.}\ \bibnamefont {Devoret}},
  \ and\ \bibinfo {author} {\bibfnamefont {I.}~\bibnamefont {Siddiqi}},\
  }\href@noop {} {\bibfield  {journal} {\bibinfo  {journal} {Review of
  Scientific Instruments}\ }\textbf {\bibinfo {volume} {80}},\ \bibinfo {eid}
  {111101} (\bibinfo {year} {2009})}\BibitemShut {NoStop}%
\bibitem [{\citenamefont {de~Groot}\ \emph {et~al.}(2010)\citenamefont
  {de~Groot}, \citenamefont {van Loo}, \citenamefont {Lisenfeld}, \citenamefont
  {Schouten}, \citenamefont {Lupascu}, \citenamefont {Harmans},\ and\
  \citenamefont {Mooij}}]{deGroot_APL_2010}%
  \BibitemOpen
  \bibfield  {author} {\bibinfo {author} {\bibfnamefont {P.~C.}\ \bibnamefont
  {de~Groot}}, \bibinfo {author} {\bibfnamefont {A.~F.}\ \bibnamefont {van
  Loo}}, \bibinfo {author} {\bibfnamefont {J.}~\bibnamefont {Lisenfeld}},
  \bibinfo {author} {\bibfnamefont {R.~N.}\ \bibnamefont {Schouten}}, \bibinfo
  {author} {\bibfnamefont {A.}~\bibnamefont {Lupascu}}, \bibinfo {author}
  {\bibfnamefont {C.~J. P.~M.}\ \bibnamefont {Harmans}}, \ and\ \bibinfo
  {author} {\bibfnamefont {J.~E.}\ \bibnamefont {Mooij}},\ }\href@noop {}
  {\bibfield  {journal} {\bibinfo  {journal} {Applied Physics Letters}\
  }\textbf {\bibinfo {volume} {96}},\ \bibinfo {eid} {123508} (\bibinfo {year}
  {2010})}\BibitemShut {NoStop}%
\bibitem [{\citenamefont {Lupascu}\ \emph {et~al.}(2007)\citenamefont
  {Lupascu}, \citenamefont {Saito}, \citenamefont {Picot}, \citenamefont
  {de~Groot}, \citenamefont {Harmans},\ and\ \citenamefont
  {Mooij}}]{Lupascu_NatPhys_2007}%
  \BibitemOpen
  \bibfield  {author} {\bibinfo {author} {\bibfnamefont {A.}~\bibnamefont
  {Lupascu}}, \bibinfo {author} {\bibfnamefont {S.}~\bibnamefont {Saito}},
  \bibinfo {author} {\bibfnamefont {T.}~\bibnamefont {Picot}}, \bibinfo
  {author} {\bibfnamefont {P.~C.}\ \bibnamefont {de~Groot}}, \bibinfo {author}
  {\bibfnamefont {C.~J. P.~M.}\ \bibnamefont {Harmans}}, \ and\ \bibinfo
  {author} {\bibfnamefont {J.~E.}\ \bibnamefont {Mooij}},\ }\href
  {http://dx.doi.org/10.1038/nphys509} {\bibfield  {journal} {\bibinfo
  {journal} {Nat Phys}\ }\textbf {\bibinfo {volume} {3}},\ \bibinfo {pages}
  {119} (\bibinfo {year} {2007})}\BibitemShut {NoStop}%
\bibitem [{\citenamefont {Schuster}\ \emph {et~al.}(2005)\citenamefont
  {Schuster}, \citenamefont {Wallraff}, \citenamefont {Blais}, \citenamefont
  {Frunzio}, \citenamefont {Huang}, \citenamefont {Majer}, \citenamefont
  {Girvin},\ and\ \citenamefont {Schoelkopf}}]{Schuster_PRL_2007}%
  \BibitemOpen
  \bibfield  {author} {\bibinfo {author} {\bibfnamefont {D.~I.}\ \bibnamefont
  {Schuster}}, \bibinfo {author} {\bibfnamefont {A.}~\bibnamefont {Wallraff}},
  \bibinfo {author} {\bibfnamefont {A.}~\bibnamefont {Blais}}, \bibinfo
  {author} {\bibfnamefont {L.}~\bibnamefont {Frunzio}}, \bibinfo {author}
  {\bibfnamefont {R.-S.}\ \bibnamefont {Huang}}, \bibinfo {author}
  {\bibfnamefont {J.}~\bibnamefont {Majer}}, \bibinfo {author} {\bibfnamefont
  {S.~M.}\ \bibnamefont {Girvin}}, \ and\ \bibinfo {author} {\bibfnamefont
  {R.~J.}\ \bibnamefont {Schoelkopf}},\ }\href {\doibase
  10.1103/PhysRevLett.94.123602} {\bibfield  {journal} {\bibinfo  {journal}
  {Phys. Rev. Lett.}\ }\textbf {\bibinfo {volume} {94}},\ \bibinfo {pages}
  {123602} (\bibinfo {year} {2005})}\BibitemShut {NoStop}%
\bibitem [{\citenamefont {Sears}\ \emph {et~al.}(2012)\citenamefont {Sears},
  \citenamefont {Petrenko}, \citenamefont {Catelani}, \citenamefont {Sun},
  \citenamefont {Paik}, \citenamefont {Kirchmair}, \citenamefont {Frunzio},
  \citenamefont {Glazman}, \citenamefont {Girvin},\ and\ \citenamefont
  {Schoelkopf}}]{Sears_dephasing_2012}%
  \BibitemOpen
  \bibfield  {author} {\bibinfo {author} {\bibfnamefont {A.~P.}\ \bibnamefont
  {Sears}}, \bibinfo {author} {\bibfnamefont {A.}~\bibnamefont {Petrenko}},
  \bibinfo {author} {\bibfnamefont {G.}~\bibnamefont {Catelani}}, \bibinfo
  {author} {\bibfnamefont {L.}~\bibnamefont {Sun}}, \bibinfo {author}
  {\bibfnamefont {H.}~\bibnamefont {Paik}}, \bibinfo {author} {\bibfnamefont
  {G.}~\bibnamefont {Kirchmair}}, \bibinfo {author} {\bibfnamefont
  {L.}~\bibnamefont {Frunzio}}, \bibinfo {author} {\bibfnamefont {L.~I.}\
  \bibnamefont {Glazman}}, \bibinfo {author} {\bibfnamefont {S.~M.}\
  \bibnamefont {Girvin}}, \ and\ \bibinfo {author} {\bibfnamefont {R.~J.}\
  \bibnamefont {Schoelkopf}},\ }\href {\doibase 10.1103/PhysRevB.86.180504}
  {\bibfield  {journal} {\bibinfo  {journal} {Phys. Rev. B}\ }\textbf {\bibinfo
  {volume} {86}},\ \bibinfo {pages} {180504(R)} (\bibinfo {year}
  {2012})}\BibitemShut {NoStop}%
\bibitem [{\citenamefont {Petrashov}\ \emph {et~al.}(1995)\citenamefont
  {Petrashov}, \citenamefont {Antonov}, \citenamefont {Delsing},\ and\
  \citenamefont {Claeson}}]{Petrashov_PRL_1995}%
  \BibitemOpen
  \bibfield  {author} {\bibinfo {author} {\bibfnamefont {V.~T.}\ \bibnamefont
  {Petrashov}}, \bibinfo {author} {\bibfnamefont {V.~N.}\ \bibnamefont
  {Antonov}}, \bibinfo {author} {\bibfnamefont {P.}~\bibnamefont {Delsing}}, \
  and\ \bibinfo {author} {\bibfnamefont {T.}~\bibnamefont {Claeson}},\ }\href
  {\doibase 10.1103/PhysRevLett.74.5268} {\bibfield  {journal} {\bibinfo
  {journal} {Phys. Rev. Lett.}\ }\textbf {\bibinfo {volume} {74}},\ \bibinfo
  {pages} {5268} (\bibinfo {year} {1995})}\BibitemShut {NoStop}%
\bibitem [{\citenamefont {Petrashov}\ \emph {et~al.}(2005)\citenamefont
  {Petrashov}, \citenamefont {Chua}, \citenamefont {Marshall}, \citenamefont
  {Shaikhaidarov},\ and\ \citenamefont {Nicholls}}]{Petrashov_PRL_2005}%
  \BibitemOpen
  \bibfield  {author} {\bibinfo {author} {\bibfnamefont {V.~T.}\ \bibnamefont
  {Petrashov}}, \bibinfo {author} {\bibfnamefont {K.~G.}\ \bibnamefont {Chua}},
  \bibinfo {author} {\bibfnamefont {K.~M.}\ \bibnamefont {Marshall}}, \bibinfo
  {author} {\bibfnamefont {R.~S.}\ \bibnamefont {Shaikhaidarov}}, \ and\
  \bibinfo {author} {\bibfnamefont {J.~T.}\ \bibnamefont {Nicholls}},\ }\href
  {\doibase 10.1103/PhysRevLett.95.147001} {\bibfield  {journal} {\bibinfo
  {journal} {Phys. Rev. Lett.}\ }\textbf {\bibinfo {volume} {95}},\ \bibinfo
  {pages} {147001} (\bibinfo {year} {2005})}\BibitemShut {NoStop}%
\bibitem [{\citenamefont {Shelly}\ \emph {et~al.}(2016)\citenamefont {Shelly},
  \citenamefont {Matrozova},\ and\ \citenamefont {Petrashov}}]{Shelly2016}%
  \BibitemOpen
  \bibfield  {author} {\bibinfo {author} {\bibfnamefont {C.~D.}\ \bibnamefont
  {Shelly}}, \bibinfo {author} {\bibfnamefont {E.~A.}\ \bibnamefont
  {Matrozova}}, \ and\ \bibinfo {author} {\bibfnamefont {V.~T.}\ \bibnamefont
  {Petrashov}},\ }\href@noop {} {\bibfield  {journal} {\bibinfo  {journal}
  {Science Advances}\ }\textbf {\bibinfo {volume} {2}},\ \bibinfo {eid}
  {e1501250} (\bibinfo {year} {2016})}\BibitemShut {NoStop}%
\bibitem [{\citenamefont {Likharev}(1986)}]{LikharevBook}%
  \BibitemOpen
  \bibfield  {author} {\bibinfo {author} {\bibfnamefont {K.~K.}\ \bibnamefont
  {Likharev}},\ }\href {https://books.google.co.uk/books?id=7lMZpE1ICUMC}
  {\emph {\bibinfo {title} {Dynamics of Josephson Junctions and Circuits}}}\
  (\bibinfo  {publisher} {Taylor \& Francis},\ \bibinfo {year}
  {1986})\BibitemShut {NoStop}%
\bibitem [{\citenamefont {Clarke}\ and\ \citenamefont
  {Braginski}(2006)}]{clarke2006squid}%
  \BibitemOpen
  \bibfield  {author} {\bibinfo {author} {\bibfnamefont {J.}~\bibnamefont
  {Clarke}}\ and\ \bibinfo {author} {\bibfnamefont {A.}~\bibnamefont
  {Braginski}},\ }\href {https://books.google.co.uk/books?id=BsTTM-nU-JkC}
  {\emph {\bibinfo {title} {The SQUID Handbook: Fundamentals and Technology of
  SQUIDs and SQUID Systems}}}\ (\bibinfo  {publisher} {Wiley},\ \bibinfo {year}
  {2006})\BibitemShut {NoStop}%
\bibitem [{\citenamefont {Kofman}\ \emph {et~al.}(2007)\citenamefont {Kofman},
  \citenamefont {Zhang}, \citenamefont {Martinis},\ and\ \citenamefont
  {Korotkov}}]{Kofman_PRB_2007}%
  \BibitemOpen
  \bibfield  {author} {\bibinfo {author} {\bibfnamefont {A.~G.}\ \bibnamefont
  {Kofman}}, \bibinfo {author} {\bibfnamefont {Q.}~\bibnamefont {Zhang}},
  \bibinfo {author} {\bibfnamefont {J.~M.}\ \bibnamefont {Martinis}}, \ and\
  \bibinfo {author} {\bibfnamefont {A.~N.}\ \bibnamefont {Korotkov}},\ }\href
  {\doibase 10.1103/PhysRevB.75.014524} {\bibfield  {journal} {\bibinfo
  {journal} {Phys. Rev. B}\ }\textbf {\bibinfo {volume} {75}},\ \bibinfo
  {pages} {014524} (\bibinfo {year} {2007})}\BibitemShut {NoStop}%
\bibitem [{\citenamefont {Gurevich}\ \emph {et~al.}(2006)\citenamefont
  {Gurevich}, \citenamefont {Kozub},\ and\ \citenamefont
  {Shelankov}}]{Gurevich2006}%
  \BibitemOpen
  \bibfield  {author} {\bibinfo {author} {\bibfnamefont {V.~L.}\ \bibnamefont
  {Gurevich}}, \bibinfo {author} {\bibfnamefont {V.~I.}\ \bibnamefont {Kozub}},
  \ and\ \bibinfo {author} {\bibfnamefont {A.~L.}\ \bibnamefont {Shelankov}},\
  }\href {https://doi.org/10.1140/epjb/e2006-00218-6} {\bibfield  {journal}
  {\bibinfo  {journal} {The European Physical Journal B - Condensed Matter and
  Complex Systems}\ }\textbf {\bibinfo {volume} {51}},\ \bibinfo {pages} {285}
  (\bibinfo {year} {2006})}\BibitemShut {NoStop}%
\bibitem [{\citenamefont {Checkley}\ \emph {et~al.}(2011)\citenamefont
  {Checkley}, \citenamefont {Iagallo}, \citenamefont {Shaikhaidarov},
  \citenamefont {Nicholls},\ and\ \citenamefont {Petrashov}}]{Checkley_2011}%
  \BibitemOpen
  \bibfield  {author} {\bibinfo {author} {\bibfnamefont {C.}~\bibnamefont
  {Checkley}}, \bibinfo {author} {\bibfnamefont {A.}~\bibnamefont {Iagallo}},
  \bibinfo {author} {\bibfnamefont {R.}~\bibnamefont {Shaikhaidarov}}, \bibinfo
  {author} {\bibfnamefont {J.~T.}\ \bibnamefont {Nicholls}}, \ and\ \bibinfo
  {author} {\bibfnamefont {V.~T.}\ \bibnamefont {Petrashov}},\ }\href {\doibase
  10.1088/0953-8984/23/13/135301} {\bibfield  {journal} {\bibinfo  {journal}
  {Journal of Physics: Condensed Matter}\ }\textbf {\bibinfo {volume} {23}},\
  \bibinfo {pages} {135301} (\bibinfo {year} {2011})}\BibitemShut {NoStop}%
\bibitem [{\citenamefont {Barends}\ \emph {et~al.}(2014)\citenamefont
  {Barends}, \citenamefont {Kelly}, \citenamefont {Megrant}, \citenamefont
  {Veitia}, \citenamefont {Sank}, \citenamefont {Jeffrey}, \citenamefont
  {White}, \citenamefont {Mutus}, \citenamefont {Fowler}, \citenamefont
  {Campbell}, \citenamefont {Chen}, \citenamefont {Chen}, \citenamefont
  {Chiaro}, \citenamefont {Dunsworth}, \citenamefont {Neill}, \citenamefont
  {O’Malley}, \citenamefont {Roushan}, \citenamefont {Vainsencher},
  \citenamefont {Wenner}, \citenamefont {Korotkov}, \citenamefont {Cleland},\
  and\ \citenamefont {Martinis}}]{Barends2014}%
  \BibitemOpen
  \bibfield  {author} {\bibinfo {author} {\bibfnamefont {R.}~\bibnamefont
  {Barends}}, \bibinfo {author} {\bibfnamefont {J.}~\bibnamefont {Kelly}},
  \bibinfo {author} {\bibfnamefont {A.}~\bibnamefont {Megrant}}, \bibinfo
  {author} {\bibfnamefont {A.}~\bibnamefont {Veitia}}, \bibinfo {author}
  {\bibfnamefont {D.}~\bibnamefont {Sank}}, \bibinfo {author} {\bibfnamefont
  {E.}~\bibnamefont {Jeffrey}}, \bibinfo {author} {\bibfnamefont {T.~C.}\
  \bibnamefont {White}}, \bibinfo {author} {\bibfnamefont {J.}~\bibnamefont
  {Mutus}}, \bibinfo {author} {\bibfnamefont {A.~G.}\ \bibnamefont {Fowler}},
  \bibinfo {author} {\bibfnamefont {B.}~\bibnamefont {Campbell}}, \bibinfo
  {author} {\bibfnamefont {Y.}~\bibnamefont {Chen}}, \bibinfo {author}
  {\bibfnamefont {Z.}~\bibnamefont {Chen}}, \bibinfo {author} {\bibfnamefont
  {B.}~\bibnamefont {Chiaro}}, \bibinfo {author} {\bibfnamefont
  {A.}~\bibnamefont {Dunsworth}}, \bibinfo {author} {\bibfnamefont
  {C.}~\bibnamefont {Neill}}, \bibinfo {author} {\bibfnamefont
  {P.}~\bibnamefont {O’Malley}}, \bibinfo {author} {\bibfnamefont
  {P.}~\bibnamefont {Roushan}}, \bibinfo {author} {\bibfnamefont
  {A.}~\bibnamefont {Vainsencher}}, \bibinfo {author} {\bibfnamefont
  {J.}~\bibnamefont {Wenner}}, \bibinfo {author} {\bibfnamefont {A.~N.}\
  \bibnamefont {Korotkov}}, \bibinfo {author} {\bibfnamefont {A.~N.}\
  \bibnamefont {Cleland}}, \ and\ \bibinfo {author} {\bibfnamefont {J.~M.}\
  \bibnamefont {Martinis}},\ }\href@noop {} {\bibfield  {journal} {\bibinfo
  {journal} {Nature}\ }\textbf {\bibinfo {volume} {508}},\ \bibinfo {pages}
  {500} (\bibinfo {year} {2014})}\BibitemShut {NoStop}%
\bibitem [{\citenamefont {Arute}\ \emph {et~al.}(2019)\citenamefont {Arute},
  \citenamefont {Arya}, \citenamefont {Babbush}, \citenamefont {Bacon},
  \citenamefont {Bardin}, \citenamefont {Barends}, \citenamefont {Biswas},
  \citenamefont {Boixo}, \citenamefont {Brandao}, \citenamefont {Buell},
  \citenamefont {Burkett}, \citenamefont {Chen}, \citenamefont {Chen},
  \citenamefont {Chiaro}, \citenamefont {Collins}, \citenamefont {Courtney},
  \citenamefont {Dunsworth}, \citenamefont {Farhi}, \citenamefont {Foxen},
  \citenamefont {Fowler}, \citenamefont {Gidney}, \citenamefont {Giustina},
  \citenamefont {Graff}, \citenamefont {Guerin}, \citenamefont {Habegger},
  \citenamefont {Harrigan}, \citenamefont {Hartmann}, \citenamefont {Ho},
  \citenamefont {Hoffmann}, \citenamefont {Huang}, \citenamefont {Humble},
  \citenamefont {Isakov}, \citenamefont {Jeffrey}, \citenamefont {Jiang},
  \citenamefont {Kafri}, \citenamefont {Kechedzhi}, \citenamefont {Kelly},
  \citenamefont {Klimov}, \citenamefont {Knysh}, \citenamefont {Korotkov},
  \citenamefont {Kostritsa}, \citenamefont {Landhuis}, \citenamefont
  {Lindmark}, \citenamefont {Lucero}, \citenamefont {Lyakh}, \citenamefont
  {Mandrà}, \citenamefont {McClean}, \citenamefont {McEwen}, \citenamefont
  {Megrant}, \citenamefont {Mi}, \citenamefont {Michielsen}, \citenamefont
  {Mohseni}, \citenamefont {Mutus}, \citenamefont {Naaman}, \citenamefont
  {Neeley}, \citenamefont {Neill}, \citenamefont {Niu}, \citenamefont {Ostby},
  \citenamefont {Petukhov}, \citenamefont {Platt}, \citenamefont {Quintana},
  \citenamefont {Rieffel}, \citenamefont {Roushan}, \citenamefont {Rubin},
  \citenamefont {Sank}, \citenamefont {Satzinger}, \citenamefont {Smelyanskiy},
  \citenamefont {Sung}, \citenamefont {Trevithick}, \citenamefont
  {Vainsencher}, \citenamefont {Villalonga}, \citenamefont {White},
  \citenamefont {Yao}, \citenamefont {Yeh}, \citenamefont {Zalcman},
  \citenamefont {Neven},\ and\ \citenamefont {Martinis}}]{Arute2019}%
  \BibitemOpen
  \bibfield  {author} {\bibinfo {author} {\bibfnamefont {F.}~\bibnamefont
  {Arute}}, \bibinfo {author} {\bibfnamefont {K.}~\bibnamefont {Arya}},
  \bibinfo {author} {\bibfnamefont {R.}~\bibnamefont {Babbush}}, \bibinfo
  {author} {\bibfnamefont {D.}~\bibnamefont {Bacon}}, \bibinfo {author}
  {\bibfnamefont {J.~C.}\ \bibnamefont {Bardin}}, \bibinfo {author}
  {\bibfnamefont {R.}~\bibnamefont {Barends}}, \bibinfo {author} {\bibfnamefont
  {R.}~\bibnamefont {Biswas}}, \bibinfo {author} {\bibfnamefont
  {S.}~\bibnamefont {Boixo}}, \bibinfo {author} {\bibfnamefont {F.~G. S.~L.}\
  \bibnamefont {Brandao}}, \bibinfo {author} {\bibfnamefont {D.~A.}\
  \bibnamefont {Buell}}, \bibinfo {author} {\bibfnamefont {B.}~\bibnamefont
  {Burkett}}, \bibinfo {author} {\bibfnamefont {Y.}~\bibnamefont {Chen}},
  \bibinfo {author} {\bibfnamefont {Z.}~\bibnamefont {Chen}}, \bibinfo {author}
  {\bibfnamefont {B.}~\bibnamefont {Chiaro}}, \bibinfo {author} {\bibfnamefont
  {R.}~\bibnamefont {Collins}}, \bibinfo {author} {\bibfnamefont
  {W.}~\bibnamefont {Courtney}}, \bibinfo {author} {\bibfnamefont
  {A.}~\bibnamefont {Dunsworth}}, \bibinfo {author} {\bibfnamefont
  {E.}~\bibnamefont {Farhi}}, \bibinfo {author} {\bibfnamefont
  {B.}~\bibnamefont {Foxen}}, \bibinfo {author} {\bibfnamefont
  {A.}~\bibnamefont {Fowler}}, \bibinfo {author} {\bibfnamefont
  {C.}~\bibnamefont {Gidney}}, \bibinfo {author} {\bibfnamefont
  {M.}~\bibnamefont {Giustina}}, \bibinfo {author} {\bibfnamefont
  {R.}~\bibnamefont {Graff}}, \bibinfo {author} {\bibfnamefont
  {K.}~\bibnamefont {Guerin}}, \bibinfo {author} {\bibfnamefont
  {S.}~\bibnamefont {Habegger}}, \bibinfo {author} {\bibfnamefont {M.~P.}\
  \bibnamefont {Harrigan}}, \bibinfo {author} {\bibfnamefont {M.~J.}\
  \bibnamefont {Hartmann}}, \bibinfo {author} {\bibfnamefont {A.}~\bibnamefont
  {Ho}}, \bibinfo {author} {\bibfnamefont {M.}~\bibnamefont {Hoffmann}},
  \bibinfo {author} {\bibfnamefont {T.}~\bibnamefont {Huang}}, \bibinfo
  {author} {\bibfnamefont {T.~S.}\ \bibnamefont {Humble}}, \bibinfo {author}
  {\bibfnamefont {S.~V.}\ \bibnamefont {Isakov}}, \bibinfo {author}
  {\bibfnamefont {E.}~\bibnamefont {Jeffrey}}, \bibinfo {author} {\bibfnamefont
  {Z.}~\bibnamefont {Jiang}}, \bibinfo {author} {\bibfnamefont
  {D.}~\bibnamefont {Kafri}}, \bibinfo {author} {\bibfnamefont
  {K.}~\bibnamefont {Kechedzhi}}, \bibinfo {author} {\bibfnamefont
  {J.}~\bibnamefont {Kelly}}, \bibinfo {author} {\bibfnamefont {P.~V.}\
  \bibnamefont {Klimov}}, \bibinfo {author} {\bibfnamefont {S.}~\bibnamefont
  {Knysh}}, \bibinfo {author} {\bibfnamefont {A.}~\bibnamefont {Korotkov}},
  \bibinfo {author} {\bibfnamefont {F.}~\bibnamefont {Kostritsa}}, \bibinfo
  {author} {\bibfnamefont {D.}~\bibnamefont {Landhuis}}, \bibinfo {author}
  {\bibfnamefont {M.}~\bibnamefont {Lindmark}}, \bibinfo {author}
  {\bibfnamefont {E.}~\bibnamefont {Lucero}}, \bibinfo {author} {\bibfnamefont
  {D.}~\bibnamefont {Lyakh}}, \bibinfo {author} {\bibfnamefont
  {S.}~\bibnamefont {Mandrà}}, \bibinfo {author} {\bibfnamefont {J.~R.}\
  \bibnamefont {McClean}}, \bibinfo {author} {\bibfnamefont {M.}~\bibnamefont
  {McEwen}}, \bibinfo {author} {\bibfnamefont {A.}~\bibnamefont {Megrant}},
  \bibinfo {author} {\bibfnamefont {X.}~\bibnamefont {Mi}}, \bibinfo {author}
  {\bibfnamefont {K.}~\bibnamefont {Michielsen}}, \bibinfo {author}
  {\bibfnamefont {M.}~\bibnamefont {Mohseni}}, \bibinfo {author} {\bibfnamefont
  {J.}~\bibnamefont {Mutus}}, \bibinfo {author} {\bibfnamefont
  {O.}~\bibnamefont {Naaman}}, \bibinfo {author} {\bibfnamefont
  {M.}~\bibnamefont {Neeley}}, \bibinfo {author} {\bibfnamefont
  {C.}~\bibnamefont {Neill}}, \bibinfo {author} {\bibfnamefont {M.~Y.}\
  \bibnamefont {Niu}}, \bibinfo {author} {\bibfnamefont {E.}~\bibnamefont
  {Ostby}}, \bibinfo {author} {\bibfnamefont {A.}~\bibnamefont {Petukhov}},
  \bibinfo {author} {\bibfnamefont {J.~C.}\ \bibnamefont {Platt}}, \bibinfo
  {author} {\bibfnamefont {C.}~\bibnamefont {Quintana}}, \bibinfo {author}
  {\bibfnamefont {E.~G.}\ \bibnamefont {Rieffel}}, \bibinfo {author}
  {\bibfnamefont {P.}~\bibnamefont {Roushan}}, \bibinfo {author} {\bibfnamefont
  {N.~C.}\ \bibnamefont {Rubin}}, \bibinfo {author} {\bibfnamefont
  {D.}~\bibnamefont {Sank}}, \bibinfo {author} {\bibfnamefont {K.~J.}\
  \bibnamefont {Satzinger}}, \bibinfo {author} {\bibfnamefont {V.}~\bibnamefont
  {Smelyanskiy}}, \bibinfo {author} {\bibfnamefont {K.~J.}\ \bibnamefont
  {Sung}}, \bibinfo {author} {\bibfnamefont {M.~D.}\ \bibnamefont
  {Trevithick}}, \bibinfo {author} {\bibfnamefont {A.}~\bibnamefont
  {Vainsencher}}, \bibinfo {author} {\bibfnamefont {B.}~\bibnamefont
  {Villalonga}}, \bibinfo {author} {\bibfnamefont {T.}~\bibnamefont {White}},
  \bibinfo {author} {\bibfnamefont {Z.~J.}\ \bibnamefont {Yao}}, \bibinfo
  {author} {\bibfnamefont {P.}~\bibnamefont {Yeh}}, \bibinfo {author}
  {\bibfnamefont {A.}~\bibnamefont {Zalcman}}, \bibinfo {author} {\bibfnamefont
  {H.}~\bibnamefont {Neven}}, \ and\ \bibinfo {author} {\bibfnamefont {J.~M.}\
  \bibnamefont {Martinis}},\ }\href@noop {} {\bibfield  {journal} {\bibinfo
  {journal} {Nature}\ }\textbf {\bibinfo {volume} {574}},\ \bibinfo {pages}
  {505} (\bibinfo {year} {2019})}\BibitemShut {NoStop}%
\bibitem [{\citenamefont {Ligato}\ \emph {et~al.}(2020)\citenamefont {Ligato},
  \citenamefont {Strambini}, \citenamefont {Paolucci},\ and\ \citenamefont
  {Giazotto}}]{ligato2020persistent}%
  \BibitemOpen
  \bibfield  {author} {\bibinfo {author} {\bibfnamefont {N.}~\bibnamefont
  {Ligato}}, \bibinfo {author} {\bibfnamefont {E.}~\bibnamefont {Strambini}},
  \bibinfo {author} {\bibfnamefont {F.}~\bibnamefont {Paolucci}}, \ and\
  \bibinfo {author} {\bibfnamefont {F.}~\bibnamefont {Giazotto}},\ }\href@noop
  {} {\enquote {\bibinfo {title} {Persistent josephson phase-slip memory with
  topological protection},}\ } (\bibinfo {year} {2020}),\ \Eprint
  {http://arxiv.org/abs/2005.14298} {arXiv:2005.14298 [cond-mat.mes-hall]}
  \BibitemShut {NoStop}%
\bibitem [{\citenamefont {Petrashov}\ \emph {et~al.}(1998)\citenamefont
  {Petrashov}, \citenamefont {Shaikhaidarov}, \citenamefont {Sosnin},
  \citenamefont {Delsing}, \citenamefont {Claeson},\ and\ \citenamefont
  {Volkov}}]{Petrashov_PRB_1998}%
  \BibitemOpen
  \bibfield  {author} {\bibinfo {author} {\bibfnamefont {V.~T.}\ \bibnamefont
  {Petrashov}}, \bibinfo {author} {\bibfnamefont {R.~S.}\ \bibnamefont
  {Shaikhaidarov}}, \bibinfo {author} {\bibfnamefont {I.~A.}\ \bibnamefont
  {Sosnin}}, \bibinfo {author} {\bibfnamefont {P.}~\bibnamefont {Delsing}},
  \bibinfo {author} {\bibfnamefont {T.}~\bibnamefont {Claeson}}, \ and\
  \bibinfo {author} {\bibfnamefont {A.}~\bibnamefont {Volkov}},\ }\href
  {\doibase 10.1103/PhysRevB.58.15088} {\bibfield  {journal} {\bibinfo
  {journal} {Phys. Rev. B}\ }\textbf {\bibinfo {volume} {58}},\ \bibinfo
  {pages} {15088} (\bibinfo {year} {1998})}\BibitemShut {NoStop}%
\end{thebibliography}%

\end{document}